\documentclass[superscriptaddress,amsmath,amssymb, aps, physrev,longbibliography,twocolumn, footinbib]{revtex4-2}

\usepackage{upgreek}
\usepackage{color}
\usepackage{graphicx}
\usepackage{nccmath}
\usepackage{bm}
\usepackage{hyperref}
\hypersetup{colorlinks,breaklinks,
            urlcolor=[rgb]{0,0,0.36},
            linkcolor=[rgb]{0,0,0.36},
            citecolor=[rgb]{0,0,0.36}}
\usepackage[mathlines]{lineno}
\usepackage{dcolumn}
\usepackage{mathtools}  
\usepackage{siunitx}    
\usepackage{dsfont}
\usepackage{extarrows}
\usepackage{epsfig,tabularx,multirow,booktabs}

\usepackage{braket}
\usepackage{chemformula}

\usepackage[normalem]{ulem}

\definecolor{cadmiumgreen}{rgb}{0.0, 0.42, 0.24} 

\usepackage{pifont} 

\usepackage{adjustbox} 
\usepackage{tabularx} 
\newcolumntype{Y}{>{\centering\arraybackslash}X} 

\newcounter{SN}

\usepackage{array}
\newcolumntype{M}[1]{>{\centering\arraybackslash}m{#1}}
\makeatletter
\renewcommand\footnotesize{%
   \@setfontsize\footnotesize\@ixpt{8}%
   \abovedisplayskip 8\p@ \@plus2\p@ \@minus4\p@
   \abovedisplayshortskip \z@ \@plus\p@
   \belowdisplayshortskip 4\p@ \@plus2\p@ \@minus2\p@
   \def\@listi{\leftmargin\leftmargini
               \topsep 4\p@ \@plus2\p@ \@minus2\p@
               \parsep 2\p@ \@plus\p@ \@minus\p@
               \itemsep \parsep}%
   \belowdisplayskip \abovedisplayskip
}
\makeatother

\newcommand{\equ}[1]
{Eq.~(\ref{#1})}
\newcommand{\figu}[1]
{Fig.~\ref{#1}}

\newif\ifreviewmode

\usepackage{xcolor}
\usepackage[normalem]{ulem}

\ifreviewmode
  
\else
  
\fi

\newcommand{\Harvard}{Department of Physics, Harvard University, Cambridge, MA 02138, USA.}

\newcommand{\ETH}{Institute for Theoretical Physics, ETH Zurich, 8093 Zurich, Switzerland.}

\newcommand{\JQI}{Joint Quantum Institute, University of Maryland and NIST, College Park, Maryland, USA.}
\newcommand{\QUICS}{Joint Center for Quantum Information and Computer Science, University of Maryland and NIST,
College Park, Maryland, USA.}

\begin{document}

\title{    Linear response across interaction regimes in two-dimensional ferromagnets
}

\author{Aaron~M\"uller}
\affiliation{\ETH}
\author{Pavel~E.~Dolgirev}
\affiliation{\JQI}
\affiliation{\QUICS}
\affiliation{\Harvard}

\author{Oleksii~Malyshev}
\affiliation{\ETH}
 
\author{Eugene~Demler}
\affiliation{\ETH}

\begin{abstract}
    Recent discoveries of two-dimensional (2D) ferromagnets have stimulated intense interest in understanding and controlling their spin transport properties. 
    A central microscopic feature of these systems is that exchange-driven magnon--magnon interactions are strongly momentum dependent: low-momentum magnons interact weakly, while high-momentum ones can scatter strongly and exhibit collective hydrodynamic behavior. 
    Understanding transport in such systems therefore requires a microscopic description capable of capturing ballistic and hydrodynamic regimes on equal footing. The natural framework is the quantum Boltzmann equation (QBE), whose solution is notoriously difficult because of the multidimensional collision integrals.
    Here, we develop a method based on an efficient representation of distribution functions as sums of Gaussians, which renders the collision integrals tractable. This approach enables accurate solution of the linearized QBE and computation of momentum- and frequency-resolved linear response in 2D ferromagnets across a broad range of temperatures and magnetic fields.
    In particular, we resolve a temperature-driven crossover from a ballistic regime dominated by weakly interacting low-momentum magnons to a collective hydrodynamic regime governed by strongly interacting high-momentum modes.
    Applying this method to monolayer CrCl$_3$, we obtain good agreement with recent nitrogen-vacancy-center dephasing experiments that reported anomalous magnetic noise consistent with magnon sound. More broadly, our work establishes a general framework for computing momentum- and frequency-resolved linear response in interacting 2D quantum systems describable within quantum Boltzmann kinetics.
\end{abstract}

 \maketitle

\section{Introduction}

The emergence of collective behavior from strong interactions is a central theme of quantum many-body physics, and the recent discovery of atomically thin ferromagnets has provided a clean and tunable setting in which to explore it~\cite{burch_magnetism_2018}.
A distinctive feature of such materials is that magnons---elementary Goldstone excitations of the ferromagnetic state---interact ``softly'' through the typically dominant, SU(2)-symmetric exchange:
a classic result of spin-wave theory is that the exchange vertex is
proportional to the scalar product of the incoming momenta,
$\mathcal{T}_{\rm ex}\propto-\bm{k}\cdot\bm{p}$~\cite{dyson_general_1956,
dyson_thermodynamic_1956, Mattis2006}.
Low-momentum magnons therefore interact only weakly, while high-momentum
magnons can scatter strongly, so that the character of
transport---ballistic or collective hydrodynamic---hinges on which modes
dominate~\cite{gurzhi_1968, halperin_hydrodynamic_1969, iacocca_perspectives_2019}.
Furthermore, unlike atoms, magnons carry no fundamentally conserved
particle number---their density is set by temperature---so that the
conserved density on which a hydrodynamic description rests is not
guaranteed {\it a priori}.

In practice, however, this condition can be met: at low temperatures and in the high-purity crystals now available~\cite{serga_yig_2010, burch_magnetism_2018}, number- and momentum-conserving exchange collisions dominate over phonon, impurity, and umklapp scattering. Materials ranging from bulk yttrium iron garnet (YIG) to atomically thin CrCl$_3$ thereby support low-dissipation magnon transport~\cite{chumak_magnon_2015, cai_atomically_2019}, a promising basis for low-power spintronics~\cite{cornelissen_long-distance_2015}; their exchange interactions are SU(2)-symmetric to percent-level accuracy---for CrCl$_3$, see the discussion below \equ{eq:HamMonolayer}. The spin-transport properties of such systems are now probed directly by
qubit magnetometers such as nitrogen-vacancy (NV) centers, which sense
the magnetic noise of the magnon gas and have reported both
diffusive~\cite{du2017control, wang_noninvasive_2022} and hydrodynamic
signatures~\cite{xue_magnon_2026}.

\begin{figure*}[t!]
\centering
\includegraphics[width=0.8\linewidth]{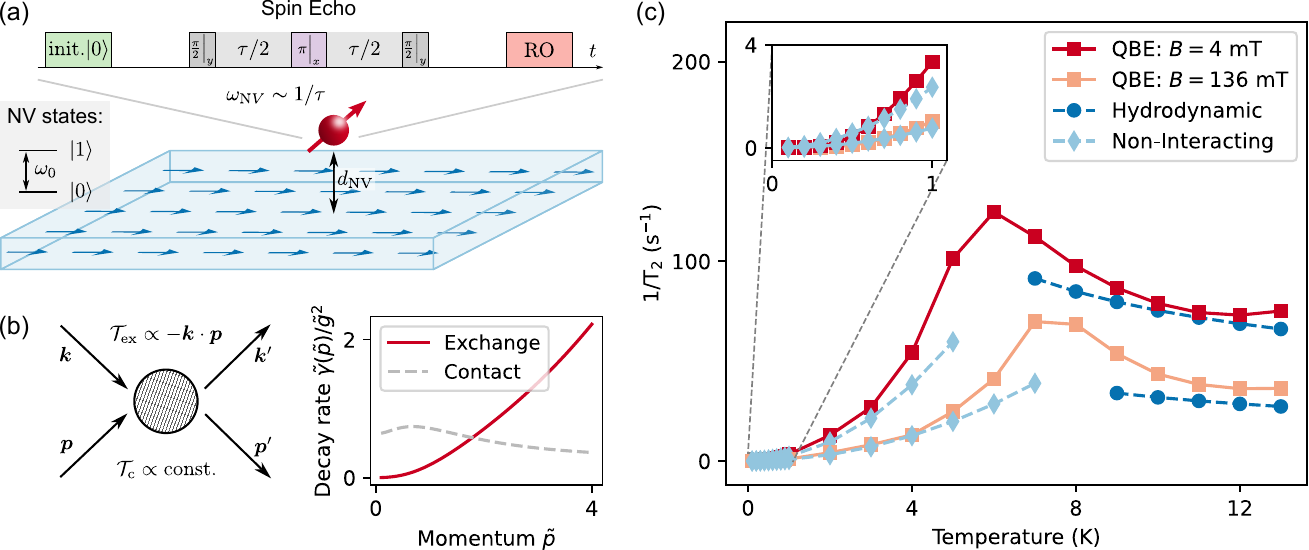}
\caption{ \textbf{Spin transport regimes of a 2D ferromagnet.} 
(a) Schematic of the setup. A spin qubit (red), such as an NV center in diamond, probes the magnetic noise from thermal spin fluctuations in the sample via spin-echo-like measurements. 
(b) A key microscopic feature of magnons---elementary spin-flip excitations on top of the ferromagnetic state---is that the exchange-driven magnon--magnon interaction vertex depends on the momenta of the scattering particles, $\mathcal{T}_{\rm ex} \propto - \bm{k}\cdot\bm{p}$ (left). As a result, high-momentum modes scatter strongly and can exhibit collective hydrodynamic behavior, whereas low-momentum modes interact only weakly.
This is reflected in the momentum-dependent magnon decay rate, which increases with $p$ (right), in contrast to the contact interaction vertex $\mathcal{T}_{\rm c} \propto \mathrm{const.}$, for which the decay rate depends only weakly on momentum.
(c) The qubit $1/T_2$ relaxation rate exhibits a nonmonotonic temperature dependence, indicating a crossover from a ballistic regime at low temperatures, where the magnon gas is dilute and effective interactions are weak, to a hydrodynamic regime at high temperatures, where strongly interacting high-momentum modes become appreciably populated and dominate the probed magnetic noise. 
This result is obtained from the full frequency- and momentum-resolved linear response computed within the linearized quantum Boltzmann equation (QBE). The crossover occurs when the exchange collision rate at the thermal momentum becomes comparable to the characteristic spatial and temporal rates probed by the NV center [\equ{eq:crossover_temperature}].
The parameters are chosen to be similar to those in Ref.~\cite{xue_magnon_2026}, which reported anomalous magnetic noise consistent with magnon hydrodynamics in atomically thin CrCl$_3$.
}
\label{fig:schematic}
\end{figure*}

The natural microscopic framework is the quantum
Boltzmann equation (QBE), which treats all momentum sectors on equal
footing and can therefore resolve whether the system is ballistic,
hydrodynamic, or in some intermediate transport regime.  
Collective behavior is encoded in the
full frequency- and momentum-resolved linear response
$\chi(\bm{q},\omega)$.
Computing it, however, is obstructed by the QBE
collision operator: a multidimensional phase-space integral, constrained by
energy and momentum conservation, that must be reevaluated across temperatures and magnetic fields and, together with the streaming terms, inverted for every probe frequency and momentum---so that an
accurate solution is, in practice, out of reach.
Prior treatments have therefore settled for one of two compromises.
Most reduce the collision integral to a single relaxation time or a
low-order variational ansatz~\cite{Basso2016, massignan2005viscous}; such approximations are inexpensive
and reproduce gross trends, but they are uncontrolled here,
where the exchange vertex is strongly momentum dependent and no single
rate describes scattering across the thermal distribution. The
alternative---retaining the full integral but evaluating it by
brute-force discretization on dense momentum grids~\cite{broido_intrinsic_2007, PhysRevB.88.045430,
cepellotti_phonon_2015, CepellottiMarzari2016}---is accurate in principle yet
becomes computationally prohibitive once the response must be mapped over the entire
$(\bm{q},\omega)$-plane and swept in temperature and magnetic field. One is thus forced to choose between accuracy and tractability.

Here, we address the collision-integral bottleneck by developing a method for solving the linearized QBE based on a Gaussian mixture model (GMM) representation of equilibrium quantum distribution functions, which renders the collision integrals tractable~\cite{dolgirev_accelerating_2024}; the formalism is presented in Sec.~\ref{sec:kinetic_theory}.
The resulting framework makes computation of the full linear response  $\chi(\bm{q},\omega)$ numerically feasible, enabling quantitative studies of bosonic many-body systems describable within the QBE kinetics, including systems with contact or exchange interactions.

As a concrete application, we study transport regimes in an exchange fluid upon increasing temperature.
At low temperatures, transport is ballistic because only a dilute
population of weakly interacting low-momentum magnons is thermally
excited.
At higher temperatures, strongly interacting high-momentum magnons
become appreciably populated, while the much more abundant low-momentum magnons remain only weakly interacting.
Although less populated, these high-momentum modes increasingly dominate the low-frequency, long-wavelength response through the emergence of a collective sound mode with linear dispersion.

This transport crossover directly manifests in the
temperature dependence of the magnetic noise probed by a qubit sensor,
such as an NV center in spin-echo measurements
[Fig.~\ref{fig:schematic}(a,c)].
At low temperatures, where transport is ballistic, the noise increases
with temperature.
As the temperature increases, high-momentum magnons drive the system
into a hydrodynamic regime.
There, a distinctive feature of exchange hydrodynamics---a consequence
of the SU(2) spin-rotation symmetry of the Heisenberg
exchange~\cite{dyson_general_1956,dyson_thermodynamic_1956,Mattis2006,rodriguez-nieva_probing_2022}---reverses
this trend, and the magnetic noise decreases with temperature.
This reversal results in the nonmonotonic profile of
Fig.~\ref{fig:schematic}(c) and offers a microscopic interpretation of
the anomalous magnetic noise recently observed in NV-center measurements
on monolayer CrCl$_3$~\cite{xue_magnon_2026}.

These results are presented in Sec.~\ref{sec:res}, where we analyze the full frequency- and momentum-resolved linear response, discuss implications for qubit spin-echo measurements, and compare exchange- and contact-interacting Bose fluids.
We conclude in Sec.~\ref{sec:conclusion} with an outlook on future directions.

\section{Theoretical framework} \label{sec:kinetic_theory}

In this section, we develop a general formalism for computing frequency- and momentum-resolved linear response in 2D many-body systems governed by the quantum Boltzmann equation (QBE) \cite{dorfman_contemporary_2021}. The main technical challenge is the evaluation of collision-integral matrix elements, which constitute the central bottleneck in solving the QBE beyond simple approximations. We show how this bottleneck can be overcome within a general framework based on an accurate representation of distribution functions as sums of Gaussians~\cite{dolgirev_accelerating_2024}, which we then apply to bosons with contact interactions as well as magnons in 2D ferromagnets with exchange-driven interactions, benchmarking the approach in the non-interacting and hydrodynamic limits.
The main physical results obtained within this formalism are presented in the following section.

\subsection{The Quantum Boltzmann Equation}
\label{subsec:QBE}

We consider an effective bosonic description of a 2D lattice model in terms of the following Hamiltonian ($\hbar = k_B = 1$):
\begin{align} \label{eqn:H_b}
    H_b  & = \sum_{\bm p} \Big(\frac{p^2}{2 m} + \Delta \Big) a^\dagger_{\bm p} a_{\bm p}  \\
    & + \frac{1}{2N} \sum_{\bm p, \bm k, \bm p', \bm k'} \mathcal T(\bm k, \bm p; \bm k', \bm p') \, \delta_{\bm k'+\bm p', \bm k + \bm p} \,  a^\dagger_{\bm k'} a^\dagger_{\bm p'} a_{\bm k} a_{\bm p}, \notag
\end{align}
where the first term describes bosons with an effective quadratic dispersion, while the second term encodes interactions through the full $T$-matrix $\mathcal T(\bm k, \bm p; \bm k', \bm p')$ for scattering processes $\{\bm k, \bm p\} \to \{\bm k', \bm p'\}$ [Fig.~\ref{fig:schematic}(b)]; here, $N$ denotes the total number of lattice sites.
Within this model, the nonequilibrium dynamics is governed by the QBE,
\begin{equation}
    \left[ \partial_t + \frac{\bm p}{m} \cdot \partial_{\bm r} - \partial_{\bm r} V_{\rm ext}(\bm r, t) \cdot \partial_{\bm p}\right] n_{\bm p}(\bm r, t) =  \mathcal I[n_{\bm p}(\bm r, t)], \label{eq:QBE}
\end{equation}
with collision integral
\begin{widetext}
\begin{align}
     \mathcal I[n_{\bm p}] = 2 {\cal A}_{\rm uc}^2 \int \frac{d^2 \bm k}{(2\pi)^2} \int \frac{d^2 \bm p'}{(2\pi)^2} \int \frac{d^2 \bm k'}{(2\pi)^2}  &|\mathcal T(\bm k, \bm p; \bm k', \bm p')|^2  (2 \pi) \delta( \varepsilon_{p} + \varepsilon_k - \varepsilon_{p'} - \varepsilon_{k'}) (2 \pi)^2 \delta(\bm k'+\bm p' - \bm k - \bm p) \notag \\
     & \times \left[n_{\bm k'}n_{\bm p'} (1 + n_{\bm k})(1 + n_{\bm p}) -  n_{\bm k}n_{\bm p}(1 + n_{\bm k'})(1 + n_{\bm p'} )\right]  , \label{eq:full_collision_integral}
\end{align}
\end{widetext}
where $\varepsilon_{p} = p^2/2m$ and ${\cal A}_{\rm uc}$ is the unit-cell area of the lattice. 
The term $V_{\rm ext}(\bm r,t)$ in Eq.~\eqref{eq:QBE} describes a weak external potential perturbation and is introduced here for later use in computing the density--density linear response function.
At thermal equilibrium, detailed balance requires $\mathcal I[n^{(0)}_{\bm p}] = 0$, which yields the Bose--Einstein distribution
\begin{align}
    n^{(0)}_{\bm p} = [z^{-1} \exp( p^2/2mT )  -1 ]^{-1}, \label{eq:n_0_eq}
\end{align}
where the fugacity $z = \exp(-\Delta/T)$ is fixed by the gap $\Delta$ and temperature $T$.

This effective bosonic model can arise microscopically from the Heisenberg Hamiltonian
\begin{align} \label{eq:HamMonolayer}
H_s   &=  - J \sum_{\langle i j\rangle } (S_i^x S_j^x + S_i^y S_j^y +   \alpha S_i^z S_j^z) \notag\\
&- g \mu_B \bm B \cdot \sum_i \bm S_i.
\end{align}
A concrete realization that we primarily focus on here is monolayer CrCl$_3$, which has a honeycomb lattice with lattice constant $a_0 \approx 5.942\,$\AA. In this material, the nearest-neighbor exchange coupling is $J\approx 10.9\,$K, and the anisotropy parameter is $\alpha \approx 0.993$, which renders CrCl$_3$ an easy-plane ferromagnet~\cite{kim_evolution_2019}. The second term in Eq.~\eqref{eq:HamMonolayer} describes the Zeeman coupling to an external magnetic field $\bm B$ (for CrCl$_3$, $g \approx 2$ and $S=3/2$).

The magnon dispersion in monolayer CrCl$_3$ has been carefully studied in Ref.~\cite{kim_evolution_2019}, while the effects of dipolar interactions were further analyzed in Ref.~\cite{xue_magnon_2026}. These works justify the parabolic dispersion used in Eq.~\eqref{eqn:H_b}, subject to two important caveats.
First, at large momenta of order the inverse lattice constant, $p \sim 1/a_0$, band-bending effects become appreciable. In the temperature regimes relevant here, $T\lesssim J$, such modes are only weakly populated, so the parabolic approximation remains reliable. If these large-momentum states were to become important, both the validity of spin-wave theory and the dominant scattering mechanisms would need to be revisited (for instance, umklapp scattering could no longer be neglected).
Second, at very small momenta the magnon dispersion is not generically parabolic either, and its precise form depends on the orientation of the applied magnetic field. For example, when the field is applied along the out-of-plane $z$ axis, the system remains U(1) rotationally symmetric, and the magnon spectrum is gapless. In this case, dipolar interactions are essential for stabilizing the long-range ferromagnetic order~\cite{xue_magnon_2026, bruno_spin_1991, Mermin1966}. 
More generally, there is a small region of momentum space near the $\Gamma$-point---whose size is controlled by the small ratio of the anisotropy and dipolar energy scales to the exchange coupling $J$---where the dispersion can deviate from the parabolic form. 
Outside this region, the dispersion is well approximated as parabolic~\cite{xue_magnon_2026}, and $\Delta$ should therefore be regarded as a pseudogap rather than a true gap.
Magnons in this momentum range dominate spin transport except at very low temperatures, where only the lowest-momentum modes with nonparabolic dispersion are appreciably populated.
In that regime, their interactions can be neglected and the computation of linear response becomes straightforward [Sec.~\ref{sec:non_interacting}].
The inverse mass is given by $m^{-1} = 3JSa^2/2$, while the pseudogap is in general set by the interplay of anisotropy, magnetic field strength and orientation, and dipolar interactions. Following Ref.~\cite{xue_magnon_2026}, in our numerical analysis we use $\Delta = 0.17\,$K for $B = 4\,$mT and $\Delta = 0.32\,$K for $B = 136\,$mT.

We now turn to the interaction vertex.
A well-known result due to Dyson~\cite{dyson_general_1956,dyson_thermodynamic_1956,Mattis2006,rodriguez-nieva_probing_2022} is that, for the Heisenberg ferromagnet with SU(2) spin-rotation symmetry, the exchange-driven magnon--magnon $T$-matrix is controlled by the momenta of the incoming scattering magnons, $\mathcal T_{\rm ex}(\bm k, \bm p; \bm k', \bm p') \propto - \bm k \cdot \bm p$ [Fig.~\ref{fig:schematic}(b)]. Motivated by this structure, we consider the following two interaction vertices:
\begin{subequations}
\begin{align}
    g_{\rm ex} \,  \bm k \cdot \bm p &\coloneqq  -\sqrt{2} {\cal A}_{\rm uc} \mathcal T_{\rm ex}(\bm k, \bm p; \bm k', \bm p') \label{eq:Tmatrix_exchange}, \\
    g_{\rm c} &\coloneqq \sqrt{2}  {\cal A}_{\rm uc}\mathcal T_{\rm c}  = \text{const.} \label{eq:Tmatrix_contact}
\end{align}
\end{subequations}
Up to a geometry-dependent numerical prefactor of order one (e.g., square versus honeycomb lattice), the coupling $g_{\rm ex}$ in Eq.~\eqref{eq:Tmatrix_exchange} scales as $g_{\rm ex} \simeq J a_0^2$.
This interaction is soft in the sense that it depends on the scalar product of the incoming momenta and therefore vanishes in the long-wavelength limit, $\bm k, \bm p \to 0$. 
This implies that scattering is kinematically suppressed at low temperatures, where only low-momentum magnons are thermally populated, and is strongly enhanced at higher temperatures, where high-momentum magnons become appreciably populated [Fig.~\ref{fig:schematic}(b,c)].
For comparison, we also consider the momentum-independent contact interaction in Eq.~\eqref{eq:Tmatrix_contact}. It is of interest in its own right, particularly in the context of atomic Bose gases with Feshbach-tunable interactions as well as excitons in atomically thin semiconductors, and can also arise from Eq.~\eqref{eq:HamMonolayer} in the presence of nonzero anisotropy, in which case $g_{\rm c} \simeq J(1-\alpha)$. 
The weak anisotropy in CrCl$_3$ implies that exchange interactions provide the dominant magnon scattering mechanism.

\subsection{Linearized Kinetic Theory}
\label{subsec:linQBE}

Our primary object of interest is the frequency- and momentum-resolved density--density response function. 
Accordingly, in the local spin frame we introduce a weak longitudinal perturbation, $-V_{\rm ext}S^z$, which couples directly to the magnon density through $S^z=S-n$:
\begin{align}
    V_{\rm ext}(\bm r,t) = V_0 e^{i\bm q\cdot \bm r - i\omega t},
\end{align}
where $V_0$ is small.
The induced change in the distribution function, to linear order in $V_0$, is parameterized as
\begin{align}
    n(\bm r,\bm p,t) \approx n^{(0)}_{\bm p} + \Delta_{\bm p} \Phi_{\bm p}(\bm q,\omega ) e^{i\bm q\cdot \bm r - i\omega t},
\end{align}
where $\Delta_{\bm p} = n^{(0)}_{\bm p}\big(1 + n^{(0)}_{\bm p}\big)$.
The resulting linearized QBE, cf. Eq.~\eqref{eq:QBE}, takes the form of an integro-differential equation for the deviations $\Phi_{\bm p}$:
\begin{widetext}
\begin{align}
    \Big(-i\omega \Delta_p   + \frac{i \bm p\cdot\bm q}{m} \Delta_p \Big)\Phi_{\bm p}  +  & \frac{i \bm p\cdot\bm q}{m T}  \Delta_p V_0   = - g_{\rm ex}^2 \int  \frac{d^2 \bm k}{(2\pi)^2} \int \frac{d^2 \bm p'}{(2\pi)^2} \int \frac{d^2 \bm k'}{(2\pi)^2}  |\bm k \cdot \bm p|^2  (2\pi) \delta( \varepsilon_{p} + \varepsilon_k - \varepsilon_{p'} - \varepsilon_{k'}
 ) \notag\\
    &\times (2\pi)^2\delta(\bm p + \bm k - \bm p' - \bm k') \times  n^{(0)}_{\bm p}n^{(0)}_{\bm k } (1 + n^{(0)}_{\bm p'}) (1 + n^{(0)}_{\bm k'}) [\Phi_{\bm p} + \Phi_{\bm k}  - \Phi_{\bm p'} - \Phi_{\bm k'} ], \label{eq:linearized_Boltzmann}
\end{align}
\end{widetext}
and an analogous expression holds for the contact interaction.
The density--density response function is then obtained as
\begin{align} \label{eqn:susz_zz}
    \chi(\bm q,\omega ) = -\frac{1}{V_0} \int \frac{d^2\bm p}{(2\pi)^2} \Delta_p \Phi_{\bm p}(\bm q,\omega ),
\end{align}
where the minus sign is introduced for notational convenience.

The right-hand side of Eq.~\eqref{eq:linearized_Boltzmann} defines a linear collision kernel acting on $\Phi_{\bm p}$. Equation~\eqref{eq:linearized_Boltzmann} can therefore be rewritten as
\begin{align}
    (\tilde{\omega}  &  - \tilde{p}\tilde{q}\cos\varphi )  \Phi_{\tilde{p}}(\varphi)  \label{eqn:LBE_v1} \\ 
     & + i\int\frac{d^2 \tilde{\bm p}' }{(2\pi)^2} {\cal C}_{\tilde{p}\tilde{p}'}(\varphi - \varphi')\Phi_{\tilde{p}'}(\varphi') = \tilde{V}_0\tilde{p}\tilde{q}\cos\varphi, \notag
\end{align}
where $\varphi$ ($\varphi'$) denotes the angle between $\tilde{\bm q}$ and $\tilde{\bm p}$ ($\tilde{\bm p}'$). Here, we introduce dimensionless variables by rescaling energies in units of $T$, e.g., $\tilde{\omega}=\omega/T$, and momenta in units of the thermal momentum $p_T=\sqrt{mT}$, e.g., $\tilde p=p/p_T$. In what follows, we suppress the tildes for notational simplicity.
An important distinction between contact and exchange interactions already appears at this stage. For the contact interaction, the dimensionless coupling $\tilde g_{\rm c}=m g_{\rm c}$ is temperature independent. By contrast, the dimensionless exchange coupling $\tilde g_{\rm ex}=m^2 T g_{\rm ex}$ grows linearly with $T$. 
This explicit temperature dependence is central to the anomalous noise behavior in Fig.~\ref{fig:schematic}(c), as varying $T$ effectively tunes the dimensionless exchange interaction strength.
The collision kernel can be decomposed as
\begin{align}
    {\cal C}_{\bm p\bm p'} = {\cal C}^{\rm I}_{\bm p\bm p'} + {\cal C}^{\rm II}_{\bm p\bm p'} + {\cal C}^{\rm III}_{\bm p\bm p'}, \notag
\end{align}
where
\begin{widetext}
\begin{subequations}\label{eq:C_decomposition}
\begin{align}
    {\cal C}^{\rm I}_{\bm p\bm p'} & = (2\pi)^2 \delta(\bm p - \bm p') \frac{\tilde{g}_{\rm ex}^2}{\Delta_p}\int \frac{d^2 \bm k}{(2\pi)^2}\int\frac{d^2 \bar{\bm k}}{(2\pi)^2} \int\frac{d^2 \bar{\bm p}}{(2\pi)^2}  |\bm k \cdot \bm p|^2
       \label{eqn:CI_gen_exc}\\
    &
    \qquad\qquad\qquad\qquad \times  (2\pi) \delta( \varepsilon_{p} + \varepsilon_k - \varepsilon_{\bar{p}} - \varepsilon_{\bar{k}}
 ) (2\pi)^2\delta(\bm p + \bm k - \bar{\bm p} - \bar{\bm k}) \, n^{(0)}_{ p }  n^{(0)}_{ k } (1 + n^{(0)}_{\bar{ p}}) (1 + n^{(0)}_{\bar{ k}}),   \notag\\
    {\cal C}^{\rm II}_{\bm p\bm p'} & = \frac{\tilde{g}_{\rm ex}^2}{\Delta_p} | \bm p \cdot \bm p'|^2 \int \frac{d^2 \bm k}{(2\pi)^2}\int\frac{d^2 \bm k'}{(2\pi)^2}   \label{eqn:C_II_gen_exc} \\
     &
    \qquad\qquad\qquad\qquad \times (2\pi) \delta( \varepsilon_{p} + \varepsilon_{p'} - \varepsilon_{k} - \varepsilon_{k'}
 )  (2\pi)^2\delta(\bm p + \bm p' - \bm k - \bm k') \,  n^{(0)}_{ p }n^{(0)}_{ p' } (1 + n^{(0)}_{ k}) (1 + n^{(0)}_{ k'}),  \notag\\
    {\cal C}^{\rm III}_{\bm p\bm p'}  & =  - \frac{2\tilde{g}_{\rm ex}^2}{\Delta_p}\int \frac{d^2 \bm k}{(2\pi)^2}\int\frac{d^2 \bm k'}{(2\pi)^2} |\bm k \cdot \bm p|^2 \label{eqn:C_III_gen_exc} \\
    & \qquad\qquad\qquad\qquad\times
      (2\pi) \delta( \varepsilon_{p} + \varepsilon_k - \varepsilon_{p'} - \varepsilon_{k'}
 ) (2\pi)^2\delta(\bm p + \bm k - \bm p' - \bm k') \,  n^{(0)}_{ p } n^{(0)}_{ k } (1 + n^{(0)}_{ p'})(1 + n^{(0)}_{ k'}) .   \notag
\end{align}
\end{subequations}
\end{widetext}
An analogous decomposition holds for the contact interaction. 
The term ${\cal C}^{\rm I}_{\bm p\bm p'} \propto \delta(\bm p - \bm p')$ and therefore contributes a singular diagonal part to the collision kernel. It is convenient to separate this contribution by writing
\begin{align}
    {\cal C}_{pp'}(\varphi - \varphi') = (2\pi)^2\delta(\bm p - \bm p') \gamma_p - {\cal M}_{pp'}(\varphi - \varphi').
\end{align}
Here, $\gamma_p$ defines nothing but the characteristic magnon decay rate, while ${\cal M}_{pp'}(\varphi-\varphi')$ denotes the remaining regular part of the collision kernel.

Exploiting rotational symmetry of the collision kernel, we expand the deviation $\Phi_{\bm p}$ in angular harmonics, 
\begin{align}
    \Phi_{ p}(\varphi) = \sum_{n} e^{in\varphi} \Phi_{p}^n.
\end{align}
In this basis, the original two-dimensional problem in Eq.~\eqref{eqn:LBE_v1} reduces to a hierarchy of coupled one-dimensional integral equations:
\begin{widetext}
\begin{align}
    (\omega + i\gamma_p)\Phi_{p}^n - \frac{p q}{2}(\Phi_{p}^{n-1} + \Phi_{p}^{n+1}) - i\int_0^\infty \frac{dp' p'}{2\pi} {\cal M}_{pp'}^{n} \,\Phi_{p'}^n = \frac{\tilde{V}_0 p q}{2}(\delta_{n,1} + \delta_{n,-1}). \label{eq:LBE_angular_harmonics}
\end{align}
\end{widetext}
The collision kernel ${\cal M}_{pp'}^{n}$ is diagonal in the angular index $n$, whereas the streaming term ($\propto q$) couples adjacent harmonics, $n \leftrightarrow n \pm 1$.
The macroscopic susceptibility is then obtained from the $n=0$ mode as
\begin{align} 
    \tilde{\chi}(\tilde{ q},\tilde{\omega} ) = -\frac{1}{\tilde{V}_0} \int_0^\infty \frac{d  \tilde{p} \, \tilde{p}}{2\pi} \Delta_{\tilde{p}} \Phi_{\tilde{p}}^0(\tilde{ q},\tilde{\omega} ). \label{eq:chi_from_QBE_dimless}
\end{align}
For future reference, $\tilde{\chi}$ obeys the scaling
\begin{align}
    \tilde \chi(\tilde q, \tilde \omega, \tilde g^2; z)=\tilde \chi(\alpha \tilde q, \alpha \tilde \omega, \alpha \tilde g^2; z),\label{eqn:rescale_chi}
\end{align}
a property that substantially facilitates our numerical analysis.

The notorious technical challenge in solving the linearized QBE~\eqref{eq:LBE_angular_harmonics} is the evaluation of the multidimensional collision kernels in Eqs.~\eqref{eq:C_decomposition}. Together with the streaming terms, the resulting kinetic operator must then be inverted over the full range of frequencies $\omega$ and momenta $q$ relevant to the response. Since the collision kernels must furthermore be reevaluated across temperatures and magnetic fields, a brute-force treatment is computationally prohibitive.
The bottleneck in evaluating Eqs.~\eqref{eq:C_decomposition} originates in the non-Gaussian character of the equilibrium distribution function $n_p^{(0)}$ in Eq.~\eqref{eq:n_0_eq}; if $n_p^{(0)}$ were Gaussian, the collision kernels could be evaluated analytically. To overcome this bottleneck, we follow Ref.~\cite{dolgirev_accelerating_2024}, which showed that $n_p^{(0)}$ can be accurately represented as a sum of a small number of Gaussians:
\begin{equation}
    n^{(0)}_p  = \Big[ \frac{1}{z} e^{p^2/2}  -1 \Big]^{-1}  \approx\sum_s a_s(z) e^{- p^2 / (2 \gamma_s^2)}.
    \label{eq:GMM}
\end{equation}
For the fugacities relevant here, an accuracy of order $10^{-10}$ is achieved with roughly ten Gaussians; the amplitudes $a_s(z)$ and variances $\gamma_s$ are taken from the \texttt{GMM-MoM} repository~\footnote{\href{https://github.com/PashaDolgirev/GMM-MoM}{\texttt{github.com/PashaDolgirev/GMM-MoM}}}. 
With this representation, the collision kernels become analytically tractable and numerically efficient; their explicit form is derived in Appendix~\ref{sup:matrix_elements}.
In practice, to solve the linearized QBE~\eqref{eq:LBE_angular_harmonics}, we truncate the angular harmonics to $|n| \leq N_\varphi$ and discretize the radial momentum $p$ on a grid of $N_p$ points. The integral equation is then reduced to a linear algebraic system of dimension $(2N_\varphi+1)N_p$,
\begin{align}
    \left[ {\cal L} - i {\cal O} \right] \bm \Phi = \bm V_0 ,
    \label{eq:matrix_equation}
\end{align}
where the matrix ${\cal L}$ is sparse,
\begin{align}
    [{\cal L}]_{pp'}^{nn'} & =   (\omega  + i \gamma_p)\delta_{p,p'} \delta_{n,n'} \notag \\
     & \quad - \frac{pq}{2} \delta_{p,p'} \big( \delta_{n,n' + 1} + \delta_{n,n' - 1}\big),
\end{align}
while ${\cal O}$ incorporates both ${\cal M}$ and the integration measure, it is dense in momentum space and therefore requires numerical evaluation.
Additional details of the numerical implementation are given in Appendix~\ref{sup:matrix_elements}, and the full codebase is available in the \texttt{LBE-GMM-solver} repository~\footnote{\href{https://github.com/mueller-aaron/LBE-GMM-solver}
{\texttt{github.com/mueller-aaron/LBE-GMM-solver}}}.

\begin{figure*}[t!]
\centering
\includegraphics[scale=0.8]{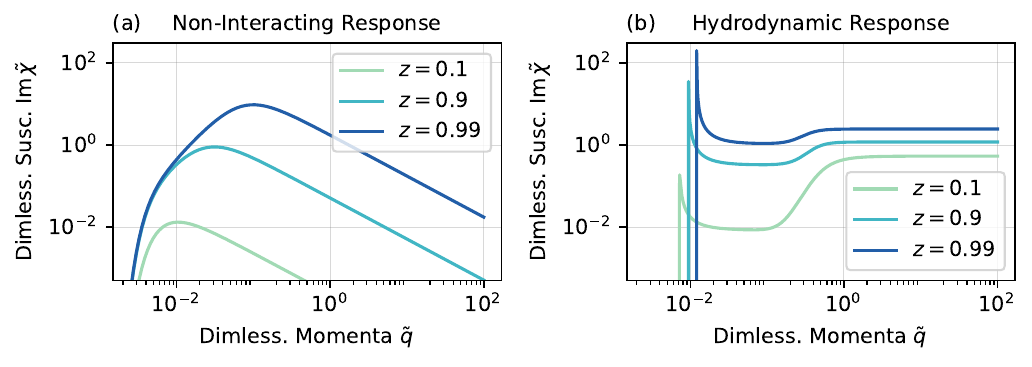} 
\caption{ 
\textbf{Non-interacting and hydrodynamic reference limits.}
Dimensionless density--density response $\mathrm{Im}\,\tilde{\chi}(\tilde q,\tilde\omega)$ as a function of dimensionless probe momentum $\tilde q$, evaluated at a fixed frequency $\tilde\omega=0.01$, in the non-interacting limit (a) and hydrodynamic limit (b).
The non-interacting response in (a) exhibits a broad incoherent single-particle continuum. 
By contrast, in the hydrodynamic limit in (b), the response develops a sharp peak at $\tilde q=\tilde\omega/\tilde c$, corresponding to the emergence of a sound mode with velocity $\tilde c$, and approaches a plateau as $\tilde q\to\infty$.
}
\label{fig:nonInt_hydro_response}
\end{figure*}

\subsection{The non-interacting Limit} \label{sec:non_interacting}

As an important benchmark for our approach to solving the QBE~\eqref{eqn:LBE_v1}, and as a physically relevant reference limit in its own right, we first analyze the non-interacting regime, in which the collision term can be neglected. The opposite limit of dominant collisions, giving rise to hydrodynamic behavior, is subsequently discussed in Sec.~\ref{subsec:Hydro}.

In the non-interacting case, Eq.~\eqref{eqn:LBE_v1} yields:
\begin{align}
    \Phi_{p}(\varphi) = V_0 \frac{p q \cos\varphi}{\omega + i \delta - p q \cos\varphi},
    \label{eq:nonint_Phi}
\end{align}
where $\delta$ is infinitesimal and introduced here to enforce causality of the retarded response function.
Projection onto the $n=0$ angular harmonic gives:
\begin{align}
    \Phi^0_p(q,\omega) = \frac{-i\omega V_0}{\sqrt{(p q)^2 - \omega^2}}.
\end{align}
This square-root singularity provides a stringent test of the solver and requires sufficiently fine radial momentum and angular grids in the numerical GMM approach outlined in Sec.~\ref{subsec:linQBE}; Figs.~\ref{fig:schematic}(c) and~\ref{fig:emergence_hydro} provide explicit benchmarks of the numerics in this non-interacting limit.

Integrating $\Phi^0_p$ over the radial momentum yields a Lindhard-like response for 2D bosons:
\begin{align}
    \text{Im} [ \tilde{\chi}(\tilde{q}, \tilde{\omega})] &=   -\frac{1}{ \tilde{V}_0} \int\frac{   d \tilde{p}}{2 \pi} \tilde{p} \Delta_{\tilde{p}}\, \text{Im} [\Phi_{\tilde{p}}^0(\tilde{q},\tilde{\omega})]  \label{eq:nonint_chi}  \\
    & = \frac{\sqrt{x_0}}{2 \pi} \int_{0}^\infty   \frac{dx}{2 \sqrt{ x }} \frac{1}{\cosh(x+y+x_0)-1},   \notag
\end{align}
where $x = \tilde{p}^2/2$, $y = \Delta/T$, and $x_0 = \tilde{\omega}^2/(2 \tilde{q}^2)$.
The resulting response, shown in Fig.~\ref{fig:nonInt_hydro_response}(a), is broad and essentially featureless, in contrast to the hydrodynamic regime we turn to discuss next.

\subsection{The Hydrodynamic Limit} \label{subsec:Hydro}

For sufficiently small probe frequency $\omega$ and momentum $q$, the system can exhibit hydrodynamic response provided that the characteristic boson--boson collision rate is large, $\omega \lesssim \gamma$, and the corresponding mean free path is short, $q l_{\rm mfp} \lesssim 1$.
In this regime, the macroscopic equations of motion follow from the conservation of (i) particle number, (ii) momentum, and (iii) kinetic energy during collision processes~\cite{kardar2007statistical,xue_magnon_2026}:
\begin{subequations}\label{eq:hydro}
\begin{gather}
    \partial_t n + \partial_\alpha (n u_\alpha) = 0, \label{eqn:hydro1}\\
    \partial_t u_\alpha + u_\beta \partial_\beta u_\alpha = -\frac{1}{m} \partial_\alpha V_{\rm ext} - \frac{1}{mn} \partial_\beta P_{\alpha\beta}, \label{eqn:hydro2}\\
    \partial_t \varepsilon + u_\alpha\partial_\alpha \varepsilon = -\frac{1}{n} \partial_\alpha h_\alpha- \frac{1}{n} P_{\alpha\beta}u_{\alpha\beta},\label{eqn:hydro3}
\end{gather}
\end{subequations}
where $u_\alpha(\bm r,t)= \langle p_\alpha/m\rangle$ is the average local stream velocity, $c_\alpha = p_\alpha/m - u_\alpha$, $P_{\alpha\beta}(\bm r,t) = n m \langle c_\alpha c_\beta \rangle$ is the pressure tensor, $\varepsilon(\bm r,t) = \langle mc^2\rangle/2$ is the local energy per particle, $h_\alpha(\bm r,t) = nm \langle c_\alpha c^2\rangle/2$ is the heat flux, and $u_{\alpha\beta} = (\partial_\alpha u_\beta + \partial_\beta u_\alpha )/2$ is the strain tensor. 
The zeroth-order hydrodynamic fields, obtained by imposing local equilibrium, are given by
\begin{gather}
    n =\frac{ m T g_1(z)}{2\pi},\qquad \varepsilon = T \frac{g_2(z)}{g_1(z)}, 
    \label{eqn:defs_hydro_0}
\end{gather}
where $g_\nu(z)$ are polylogarithm functions.
At first order away from local equilibrium, dissipative corrections enter through the shear viscosity $\mu$ and thermal conductivity $\varkappa$ as~\cite{kardar2007statistical, dorfman_contemporary_2021}:
\begin{subequations}\label{eqn:main_form_hydro}
\begin{gather}
    P_{\alpha\beta} = n \varepsilon\delta_{\alpha\beta} - \mu (\partial_\alpha u_\beta + \partial_\beta u_\alpha - \delta_{\alpha \beta} \partial_\gamma u_\gamma),\\
    h_\alpha =  - \varkappa \partial_\alpha T, 
\end{gather}
\end{subequations}
Although the transport coefficients $\mu$ and $\varkappa$ can be estimated microscopically from the QBE~\eqref{eq:QBE}, as detailed in Appendix~\ref{sec:microscopic_transport_coeff}, we treat them for now as phenomenological parameters.

Linearizing Eqs.~\eqref{eq:hydro}-\eqref{eqn:main_form_hydro} with respect to the external field $V_{\rm ext}$ yields the hydrodynamic density--density response function~\cite{xue_magnon_2026}:
\begin{align}
    \chi(q, \omega) = - \frac{n^2 q^2 }{m n \omega^2 + i \mu \omega q^2 -  n q^2 \displaystyle \frac{2 \varepsilon n \omega  + i c_3  T \varkappa q^2 }{n \omega  + i c_1 \varkappa q^2}}, \label{eq:chi_from_hydro}
\end{align}
where $c_1 =  g_1(z) g_1'(z) / (2 g_2(z) g_1'(z) - g_2'(z) g_1(z))$ and $c_3 =  g_1(z) g_2'(z)/(2 g_2(z) g_1'(z) - g_2'(z) g_1(z))$.
A representative profile is shown in Fig.~\ref{fig:nonInt_hydro_response}(b), where the response develops a sharp peak associated with the emergence of a propagating sound mode.

A natural question is then whether the mode propagates in the isothermal or isentropic regime.
This distinction is controlled by the thermal conductivity $\varkappa$, which determines the relaxation rate of temperature fluctuations relative to the timescale of the probe.
To make this explicit, we consider two limiting cases.

In the isentropic limit, $\varkappa = 0$, corresponding to slowly relaxing thermal fluctuations, the imaginary part of the susceptibility takes the Lorentzian form
\begin{align}
    \text{Im}[\chi_S(\omega,q)] =  \frac{\mu}{m^2} \frac{\omega q^4}{(\omega^2 - c_S^2 q^2)^2 + (\mu \omega q^2/mn)^2}, \label{eq:ImChi_isentropic}
\end{align}
where 
\begin{align}
    c_S^2 = \frac{2\varepsilon}{m} \eqqcolon \frac{T}{m}\tilde c_S^2
\end{align}
is the Laplace (isentropic) speed of sound.
In the opposite isothermal limit, $\varkappa\to\infty$, corresponding to rapidly relaxing thermal fluctuations, Eq.~\eqref{eq:ImChi_isentropic} retains the same form, with the Laplace sound velocity replaced by the Newton (isothermal) sound velocity
\begin{align}
c_T^2
=
\frac{T c_3(z)}{m c_1(z)}
\eqqcolon
\frac{T}{m}\tilde c_T^2,
\qquad
\tilde c_T^2
=
\frac{g'_2(z)}{g'_1(z)}.
\label{eq:isothermal_speed_of_sound}
\end{align}

The transport coefficients therefore play distinct roles: the thermal conductivity $\varkappa$ determines whether the sound mode propagates in the isentropic or isothermal regime, while the viscosity $\mu$ sets the linewidth of the sound resonance.

Finally, Fig.~\ref{fig:schematic}(c) demonstrates quantitative agreement between the QBE solver and the hydrodynamic prediction of Eq.~\eqref{eq:chi_from_hydro} at high temperatures, where the effective dynamics are expected to become hydrodynamic. For the hydrodynamic response, the transport coefficients were estimated microscopically as described in Appendix~\ref{sec:microscopic_transport_coeff}.

\section{Results and Discussion} 
\label{sec:res}

\begin{figure*}[!htb]
\centering
\includegraphics[scale=0.7]{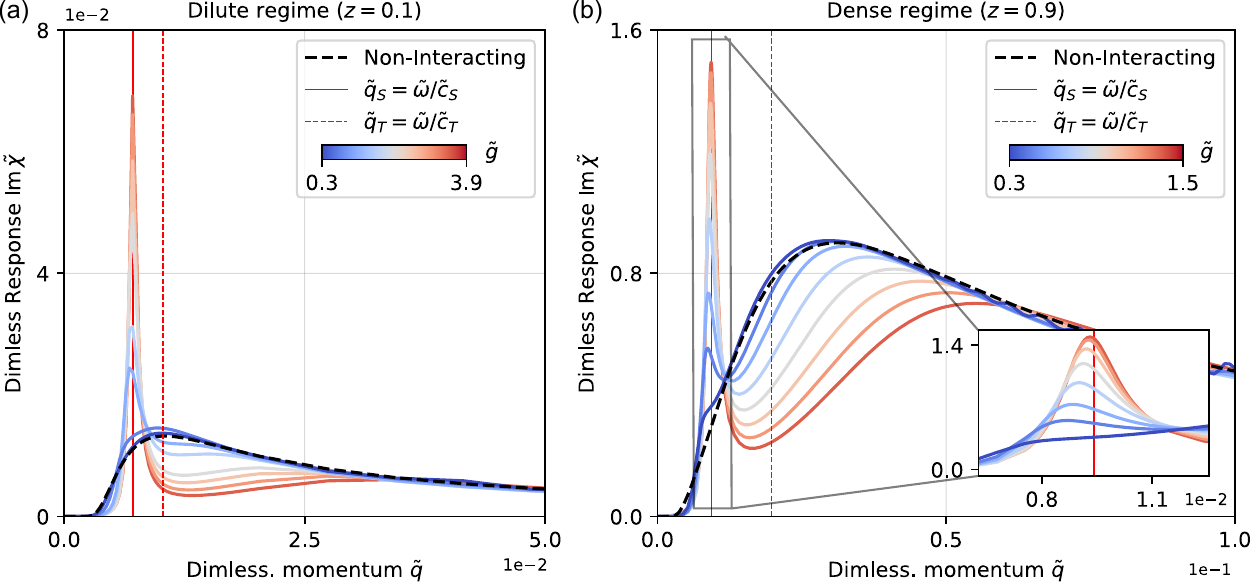} 
\caption{
\textbf{Magnon transport} in the dilute ($z=0.1$) (a) and dense ($z=0.9$) (b) regimes. 
The density--density linear response $\mathrm{Im}\,\tilde{\chi}(\tilde{\omega},\tilde{q})$, computed by solving the full QBE, is shown as a function of $\tilde q$ at the fixed frequency $\tilde\omega=0.01$.
As the dimensionless interaction strength $\tilde g$ increases, the response evolves from a broad, featureless ballistic spectrum to a spectrum with a sharp collective sound mode centered near the isentropic sound momentum, $\tilde q_S=\tilde\omega/\tilde c_S$. The inset in (b) highlights the emergence of the sound peak and its convergence toward the isentropic prediction $\tilde q_S$.
}
\label{fig:emergence_hydro}
\end{figure*}

\subsection{Linear response across magnon transport regimes}

\begin{figure*}[htb!]
\centering
\includegraphics[width=0.7\linewidth]{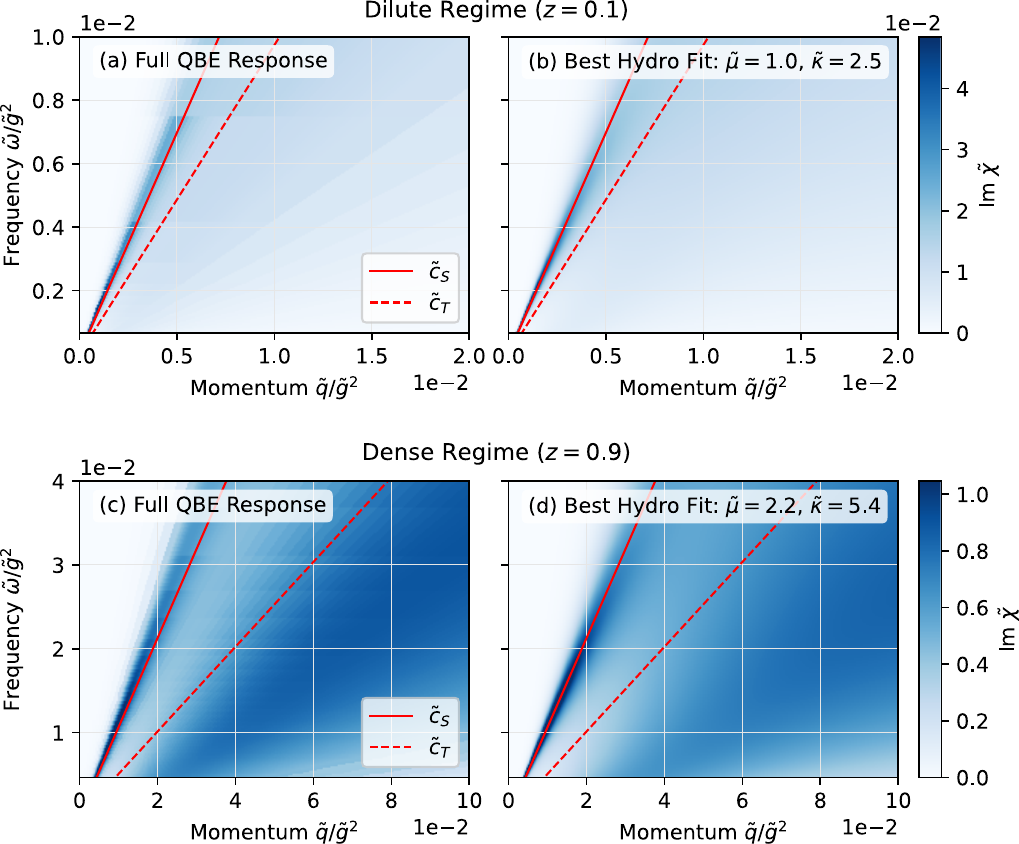} 
\caption{
\textbf{Full frequency- and momentum-resolved linear response of a 2D ferromagnet} in dilute (top row, $z = 0.1$) and dense (bottom row, $z = 0.9$) regimes.
Panels (a) and (c) show the full QBE solution for $\mathrm{Im}\,\tilde{\chi}(\tilde{\omega},\tilde{q})$, reconstructed from the data in Fig.~\ref{fig:emergence_hydro} using the scaling relation~\eqref{eqn:rescale_chi}.
Panels (b) and (d) show the corresponding best fits of the hydrodynamic response~\eqref{eq:chi_from_hydro} to the QBE solution, with $\tilde\mu$ and $\tilde\varkappa$ as the only fitting parameters [see Appendix~\ref{sup:fitting_details} for details]. 
The fitted response well captures the QBE one, including the emergence of the sound mode and the enhanced dissipation in the dense regime.
}
\label{fig:Combined_2D_hydro_map}
\end{figure*}

We first present the full linear response of a 2D ferromagnet obtained by solving the QBE.
Figure~\ref{fig:emergence_hydro} shows the evolution of the density--density response function $\mathrm{Im}\,\tilde{\chi}(\tilde{\omega},\tilde{q})$ with increasing dimensionless exchange interaction strength $\tilde g_{\rm ex}$. 
In the weakly interacting regime, the spectrum exhibits a broad, featureless response characteristic of ballistic transport [cf. Fig.~\ref{fig:nonInt_hydro_response}(a)]. 
As the interaction strength increases, a sharp collective sound resonance emerges, signaling the onset of hydrodynamic behavior [cf. Fig.~\ref{fig:nonInt_hydro_response}(b)]. 
Importantly, this sound resonance develops near the isentropic, rather than isothermal, sound momentum.
Figure~\ref{fig:Combined_2D_hydro_map} extends this analysis to the full frequency- and momentum-resolved spectrum, reconstructed from Fig.~\ref{fig:emergence_hydro} using the scaling relation~\eqref{eqn:rescale_chi}. The spectrum clearly exhibits a sharp peak following the linear isentropic sound-mode dispersion.

Throughout Figs.~\ref{fig:emergence_hydro} and~\ref{fig:Combined_2D_hydro_map}, we compare the dilute ($z=0.1$) and dense ($z=0.9$) regimes of the magnon gas. 
The dilute regime provides an important numerical benchmark: in this limit, the Gaussian expansion in Eq.~\eqref{eq:GMM} reduces to a single Gaussian of Maxwell--Boltzmann form, allowing us to further validate the numerical implementation of the collision integral.
More importantly, the two regimes reveal how exchange interactions govern magnon transport.
In the dilute regime, only low-momentum magnons are appreciably populated, implying that their exchange interaction, $\propto -\bm{k}\cdot\bm{p}$, is expected to be kinematically suppressed, so that the onset of collective behavior requires a larger interaction strength.
By contrast, in the dense regime, high-momentum magnons acquire an appreciable thermal population. 
These magnons interact strongly and can readily form a collective hydrodynamic flow, whereas the much more abundant low-momentum magnons remain only weakly interacting.
The resulting magnon transport therefore reflects a competition between a low population of strongly interacting high-momentum magnons and an abundant population of weakly interacting low-momentum magnons. 
Although both regimes exhibit the same qualitative ballistic-to-hydrodynamic crossover upon increasing the interaction strength [Fig.~\ref{fig:emergence_hydro}], this competition produces substantial quantitative differences. 
Resolving this competition is precisely what necessitates the full QBE framework developed here.

Finally, the full QBE spectral maps in Fig.~\ref{fig:Combined_2D_hydro_map}(a,c) are used to extract the magnon viscosity $\tilde{\mu}_{\rm ex}$ and thermal conductivity $\tilde\varkappa_{\rm ex}$ by fitting the analytical hydrodynamic response in Eq.~\eqref{eq:chi_from_hydro} to the QBE solution.
As detailed in Appendix~\ref{sup:fitting_details}, this yields $\tilde{\mu}_{\rm ex}=1.0$ and $\tilde\varkappa_{\rm ex}=2.5$ in the dilute regime, and $\tilde{\mu}_{\rm ex}=2.2$ and $\tilde\varkappa_{\rm ex}=5.4$ in the dense regime.
The corresponding hydrodynamic responses accurately capture the full QBE results, as shown in Fig.~\ref{fig:Combined_2D_hydro_map}(b,d).
The more than twofold enhancement of the extracted transport coefficients in the dense regime reflects the accumulation of low-momentum magnons with strongly suppressed scattering rates, which, despite their large occupation, participate only weakly in the collective hydrodynamic flow.

\subsection{    Qubit dephasing across magnon transport regimes }

\begin{figure*}[t!]
\centering
\includegraphics[scale=0.7]{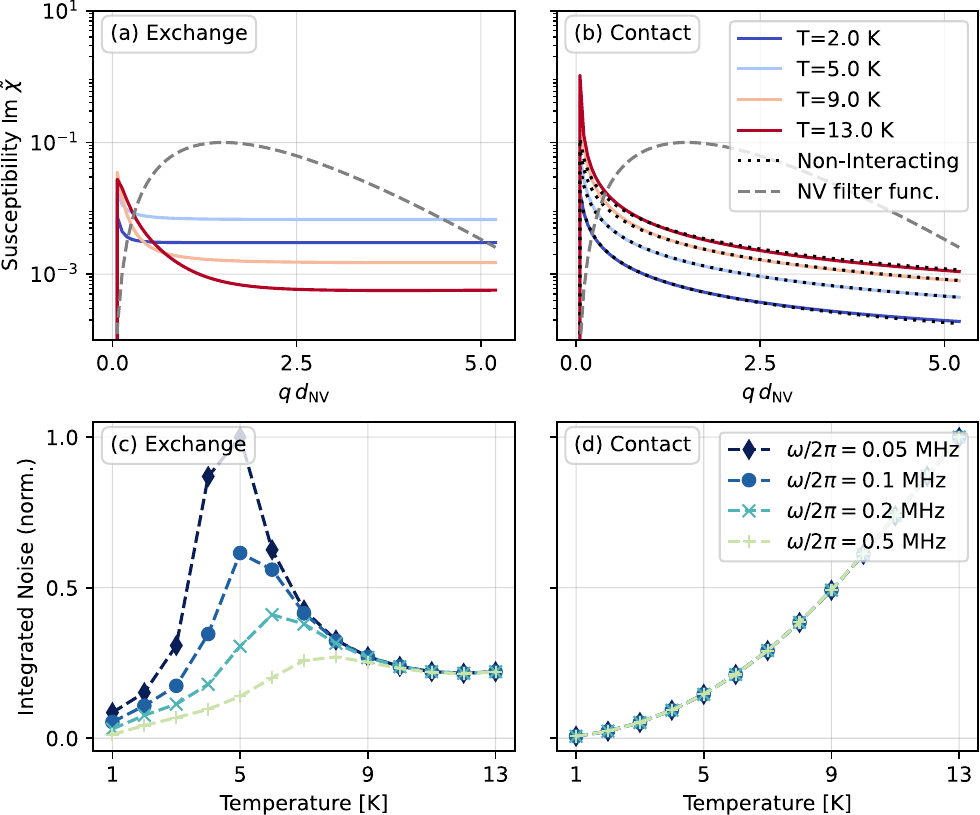}
\caption{
\textbf{Qubit signatures of exchange and contact Bose fluids.}
(a,b) Temperature evolution of the dimensionless momentum-resolved susceptibility
$\mathrm{Im}\,\tilde{\chi}(q,\omega)$ evaluated at
$\omega/2\pi=0.1\,$MHz and $B=4\,$mT for
(a) exchange and (b) contact fluids.
The dashed gray curves represent the qubit momentum filtering function
[Eq.~\eqref{eq:noise_susceptibility_relation}].
(a) In the exchange fluid, this filtering function predominantly weights high momenta, where the susceptibility approaches a viscosity-dependent plateau [Eq.~\eqref{eq:chi_from_hydro_q_infty_limit2}].
(b) The contact fluid likewise exhibits a pronounced low-$q$ sound peak. Unlike in the exchange fluid, however, the high-momentum tail, rather than saturating, decays rapidly and merges with the non-interacting susceptibility (dotted lines).
Thus, the qubit predominantly probes the non-interacting high-momentum response rather than the collective hydrodynamic one.
(c, d) Normalized integrated qubit noise as a function of temperature for various probe frequencies. 
(c) The exchange fluid exhibits a crossover from a ballistic regime, in which the qubit noise increases with increasing temperature, to an anomalous viscosity-driven hydrodynamic regime, in which it instead decreases. 
The crossover is progressively suppressed with increasing probe frequency and becomes barely resolvable for $\omega/2\pi\gtrsim1\,\mathrm{MHz}$.
(d) By contrast, the integrated qubit noise of the contact fluid increases monotonically with temperature. The curves for different probe frequencies collapse onto a single curve owing to the exact frequency cancellation of the non-interacting high-momentum response [see main text].
}
\label{fig:Tdep_integrated_response_contact_vs_exchange}
\end{figure*}

We next discuss that the ballistic-to-hydrodynamic crossover can be directly detected using NV-center spin-echo experiments [Fig.~\ref{fig:schematic}(a,c)]. 
At low temperatures, the dilute magnon gas is expected to exhibit ballistic transport, resulting in a qubit noise that increases monotonically with temperature. 
As the temperature increases, high-momentum, strongly interacting magnons become thermally populated, driving a crossover to the hydrodynamic regime at low frequencies $\omega\lesssim\gamma$ and low momenta satisfying $q l_{\rm mfp}\lesssim 1$. 
Spin-echo measurements naturally probe this low-frequency regime and, as we show below, access the momentum window where hydrodynamic behavior is expected to emerge.

Such measurements probe frequencies $\omega_{\rm NV}\sim1/\tau\lesssim1\,$MHz, several orders of magnitude below the magnon gap, which typically lies in the GHz range, making $\omega_{\rm NV}$ the lowest energy scale in the problem. As such, direct single-magnon excitations are energetically inaccessible, and the measured magnetic noise is instead dominated by longitudinal spin fluctuations, which encode the density fluctuations of the magnon fluid. The measured spin-echo signal,
$\mathcal C(\tau)=e^{-\Phi(\tau)},$
is related to the magnetic noise spectrum $\mathcal N(\omega)$ arising from magnetic fluctuations in the 2D ferromagnet through~\cite{machado_quantum_2022,xue_magnon_2026,PhysRevLett.132.246504}:
\begin{align}
    \Phi(\tau)=4\gamma_{\rm NV}^2\int\frac{d\omega}{2\pi}
    {\cal W}_\tau(\omega)\,\mathcal N(\omega),
    \label{eq:phi_noise_relation}
\end{align}
where $\gamma_{\rm NV}$ is the NV gyromagnetic ratio and
${\cal W}_\tau(\omega)=\omega^{-2}\sin^4(\omega\tau/4)$ is the spin-echo frequency filter function sharply peaked at $\omega_{\rm NV}\equiv1/\tau$.

The magnetic noise arises from the dipolar coupling between the qubit spin and the ferromagnet and can be related to the density--density response function as~\cite{machado_quantum_2022,xue_magnon_2026,PhysRevLett.132.246504}
\begin{align}
    \mathcal N(\omega)\propto
    \int_0^\infty\frac{dq}{2\pi}
    e^{-2qd_{\rm NV}}q^3
    \coth\!\left(\frac{\omega}{2T}\right)
    \mathrm{Im}\,\chi(q,\omega).
    \label{eq:noise_susceptibility_relation}
\end{align}
The factor $e^{-2qd_{\rm NV}}q^3$ acts as a momentum filtering function peaked at $q_{\rm NV}\simeq3/2d_{\rm NV}$ [Fig.~\ref{fig:Tdep_integrated_response_contact_vs_exchange}(a,b)]. 
For monolayer CrCl$_3$, a realistic qubit--sample separation is $d_{\rm NV}\simeq65\,$nm, while simple estimates place the magnon mean free path on the order of a few nm at $T\simeq10\,$K~\cite{xue_magnon_2026}, placing the experimentally relevant momentum window within the hydrodynamic regime.
Combining Eqs.~\eqref{eq:phi_noise_relation} and~\eqref{eq:noise_susceptibility_relation} and defining the dephasing rate through $1/T_2=2\Phi(\tau)/\tau$, we obtain
\begin{align}
    \frac{1}{T_2} =  \frac{2C}{d_{\rm NV}^4 } & \,\int \frac{d \bar\omega}{2 \pi} \frac{ \sin^4 (\bar \omega/4)}{\bar \omega^2} \coth\Big( \frac{\bar\omega}{2\tau T}\Big) \notag \\ 
    &\times \int_0^\infty \frac{  d  \bar q}{2\pi} e^{-2 \bar q } \,\bar q^3 \, \text{Im}\Big[ \chi\Big(\frac{\bar q}{d_{\rm NV}},\frac{\bar \omega}{\tau}\Big)\Big], \label{eq:T2_equlibrium_noise}
\end{align}
where $C$ is a dimensionful prefactor determined by the qubit--sample dipolar coupling and the orientations of the NV quantization axis and the equilibrium magnetization~\footnote{Following Ref.~\cite{xue_magnon_2026}, $C(\vartheta_{\rm NV},\vartheta_0)= (\mu_0\gamma_S\gamma_{\rm NV})^2(3+\cos^2\vartheta_0+\cos^2\vartheta_{\rm NV}+3\cos^2\vartheta_0\cos^2\vartheta_{\rm NV})/4$, where $\vartheta_{\rm NV}$ and $\vartheta_0$ denote the angles of the NV quantization axis and the equilibrium magnetization relative to the $z$ axis. 
In our numerical estimates, we use $\vartheta_{\rm NV}=54^\circ$ at 136\,mT and $\vartheta_{\rm NV}=0^\circ$ at 4\,mT, while $\vartheta_0\approx90^\circ$ for both fields.
}, and $\bar\omega=\omega\tau$ and $\bar q=qd_{\rm NV}$ are dimensionless integration variables.

Figure~\ref{fig:schematic}(c) shows the temperature dependence of the QBE dephasing rate $1/T_2$ for two representative magnetic fields using parameters comparable to those of monolayer CrCl$_3$~\cite{xue_magnon_2026} ($d_{\rm NV} = 65\,$nm and $\tau=5\,\upmu$~\footnote{To facilitate convergence of the QBE solver, we use an interaction strength of $J=4.4\,$K, which is slightly lower than that of monolayer CrCl$_3$.}). 
The resulting non-monotonic temperature dependence provides a direct signature of the ballistic-to-hydrodynamic crossover.
At low temperatures, $T\lesssim5\,$K, the magnetic noise increases monotonically with temperature, consistent with the non-interacting ballistic regime [light blue diamonds].
As the temperature increases, high-momentum, strongly interacting magnons become appreciably populated and progressively dominate the magnetic noise.
As such, the response crosses over from the ballistic regime to one governed by collective hydrodynamic flow, and at higher temperatures, $T\gtrsim8\,$K, the dephasing rate decreases monotonically, as expected for exchange-driven hydrodynamic transport [dark blue circles, computed using Eq.~\eqref{eq:T2_time_isothermal_hydro}].
We remark that in the experiment of Ref.~\cite{xue_magnon_2026} only the anomalous high-temperature trend was observed because the lower-temperature ballistic regime was not experimentally accessible.
Nevertheless, the dephasing rates in Fig.~\ref{fig:schematic}(c) are comparable in magnitude to the measured values, providing a natural interpretation of the experiment. 
An exact quantitative comparison is presently not possible, however, partially because the computed signal depends sensitively on the sample--probe distance $d_{\rm NV}$ [Eq.~\eqref{eq:T2_time_isothermal_hydro}], scaling as $d_{\rm NV}^{-4}$. As such, even a small uncertainty in $d_{\rm NV}$ can have a pronounced effect on the predicted dephasing rate, as well as on the extracted magnon viscosity $\mu_{\rm ex}$ and sound velocity $c_T$.
Moreover, our simplified theory neglects the possible temperature dependence of both the pseudogap $\Delta$ and magnon mass $m$, which may further complicate direct comparisons with the experiment.

The anomalous high-temperature trend can be understood as follows. Figure~\ref{fig:Tdep_integrated_response_contact_vs_exchange}(a) shows the temperature evolution of the momentum-resolved exchange susceptibility $\mathrm{Im}\,\tilde{\chi}(q,\omega)$. Two characteristic features emerge: a low-$q$ peak associated with the development of a collective sound mode and a momentum-independent plateau at large $q$ [cf. Fig.~\ref{fig:nonInt_hydro_response}(b)]. 
Crucially, the qubit momentum-filtering function places most of its weight on this high-momentum plateau, where the susceptibility in Eq.~\eqref{eq:chi_from_hydro} reduces to
\begin{align}
    \mathrm{Im}\,\chi(q\to\infty,\omega)
    \simeq
    \frac{\mu_{\rm ex}\omega}{m^2c_T^4}.
    \label{eq:chi_from_hydro_q_infty_limit2}
\end{align}
Interestingly, the measured signal is therefore insensitive to the isentropic sound mode and thermal conductivity, and is instead governed solely by the isothermal sound velocity $c_T$ and the magnon viscosity $\mu_{\rm ex}$. 
Combining Eqs.~\eqref{eq:T2_equlibrium_noise} and~\eqref{eq:chi_from_hydro_q_infty_limit2}, we obtain the direct relation between the dephasing rate and the viscosity~\cite{xue_magnon_2026} ($T \gg \omega_{\rm NV}$):
\begin{align}
    \frac{1}{T_2}
    =
    \frac{3C}{64\pi d_{\rm NV}^4}
    \frac{\mu T}{\Delta^2\ln^2(\Delta/T)}.
    \label{eq:T2_time_isothermal_hydro}
\end{align}
The defining signature of the exchange fluid is the scaling $\mu_{\rm ex}\sim1/T$~\cite{dyson_general_1956}. This scaling compensates the explicit factor of $T$ in Eq.~\eqref{eq:T2_time_isothermal_hydro}, leaving only a weak residual temperature dependence and making the dephasing rate decrease with increasing temperature, thereby explaining the anomalous high-temperature trend in Fig.~\ref{fig:schematic}(c) and~\figu{fig:Tdep_integrated_response_contact_vs_exchange}(c).

Three remarks are in order. First, the $1/T_2$ dephasing rate is systematically larger for the low-field configuration ($B=4\,$mT) than for the high-field one ($B=136\,$mT) [Fig.~\ref{fig:schematic}(c)]. This is consistent with the isothermal prediction of Eq.~\eqref{eq:T2_time_isothermal_hydro}, as the larger pseudogap for $136\,$mT suppresses the thermal magnon population and thereby reduces the overall magnetic noise. 
Second, the ballistic-to-hydrodynamic crossover occurs around $T_\times=6$--$7\,$K, depending on the magnetic field. 
This temperature can be estimated by comparing the exchange decay rate at the thermal momentum $p_T=\sqrt{mT}$ with the characteristic rates set by the spatial and temporal scales probed by the NV center [see Appendix~\ref{sec:crossover_estimate}].
We define the dimensionless ratio
\begin{align}
   {\cal R}_T(T;q,\omega)
   =\frac{\gamma_{\rm ex}(p_T,T)}
   {\sqrt{q^2T/m+\omega^2}}, \label{eq:crossover_temperature}
\end{align}
such that the crossover is expected parametrically at ${\cal R}_T\sim1$.
For the crossover temperature $T_\times\simeq6$--$7\,$K in Fig.~\ref{fig:schematic}(c), and evaluated at the frequency and momentum scales of the NV probe, we find ${\cal R}_\times\equiv{\cal R}_T(T_\times;q_{\rm NV},\omega_{\rm NV})\simeq0.3$, separating the deep hydrodynamic regime, ${\cal R}_T\gg1$, from the ballistic regime, ${\cal R}_T\ll1$.
Below this temperature, collisions are no longer sufficiently frequent to maintain local equilibrium over the sensing volume.
Finally, the non-monotonic temperature dependence itself depends on the probe frequency, as shown in Fig.~\ref{fig:Tdep_integrated_response_contact_vs_exchange}(c). The low-temperature increase persists up to a crossover temperature that shifts to higher values with increasing frequency. For $\omega/2\pi\gtrsim1\,$MHz, the probing timescale becomes shorter than the collision time, $1/\omega<1/\gamma$, preventing the formation of collective hydrodynamic response and thereby suppressing the anomalous high-temperature behavior.

\subsection{Qubit Signatures of a Contact Bose Fluid}

We finally consider a Bose fluid with contact interactions and contrast its qubit response with that of the exchange fluid discussed above. The key distinction is that the contact interaction vertex is momentum independent and therefore does not exhibit the low-momentum kinematic bottleneck characteristic of exchange interactions.

Compared with the exchange fluid, the momentum-resolved spectral function $\mathrm{Im}\,\tilde{\chi}(q,\omega)$ also exhibits a pronounced low-$q$ peak associated with the emergence of a collective sound mode [Fig.~\ref{fig:Tdep_integrated_response_contact_vs_exchange}(b)]. Remarkably, this resonance is nearly two orders of magnitude stronger than the corresponding exchange resonance [Fig.~\ref{fig:Tdep_integrated_response_contact_vs_exchange}(a)] for the same dimensionless interaction strength.

Unlike the exchange fluid, however, the high-$q$ tail of the contact susceptibility, rather than saturating to a plateau, rapidly decays and merges with the non-interacting ballistic response [dotted lines in Fig.~\ref{fig:Tdep_integrated_response_contact_vs_exchange}(b)]. 
Since the NV momentum-filtering function weights this high-momentum tail, the qubit essentially probes the non-interacting ballistic regime.
Notably, the responses for all probed frequencies collapse onto a single curve as shown in Fig.~\ref{fig:Tdep_integrated_response_contact_vs_exchange}(d).
In this low-frequency limit ($\omega \ll T$), expanding the phase velocity term ($x_0 \ll 1$) in \equ{eq:nonint_chi} shows that the ballistic susceptibility scales linearly with frequency, $\text{Im}\, \chi \propto \omega$. 
This linear dependence is then canceled out by the thermal prefactor, $\coth(\omega/2T) \approx 2T/\omega$, rendering the integrated noise frequency independent.
Upon further cooling, the finite magnon gap $\Delta$ thermally freezes out the magnons, and the noise becomes exponentially suppressed.

The resulting temperature dependence is therefore monotonic and markedly distinct from that of the exchange fluid. We thus conclude that the anomalous high-temperature behavior of the dephasing profile in Fig.~\ref{fig:schematic}(c) and Fig.~\ref{fig:Tdep_integrated_response_contact_vs_exchange}(c) is a direct fingerprint of exchange-driven magnon hydrodynamics.

\section{Conclusion and Outlook}
\label{sec:conclusion}

We have developed a general framework for computing frequency- and momentum-resolved linear response in 2D many-body systems governed by the QBE. The central technical challenge---the numerical evaluation of multidimensional collision-integral matrix elements---is overcome for systems with parabolic dispersion by representing the Bose--Einstein distribution as a sum of a few Gaussians~\cite{dolgirev_accelerating_2024}, an approximation accurate across all physical regimes of interest that renders the collision integrals tractable and efficient to evaluate.

The main physical result is the computation of the full density--density response $\chi(q,\omega)$ of 2D bosonic systems with either contact or exchange-driven interactions [see Figs.~\ref{fig:Combined_2D_hydro_map} and~\ref{fig:Tdep_integrated_response_contact_vs_exchange}].
The exchange vertex ${\cal T}_{\rm ex}(\bm k, \bm p; \bm k', \bm p') \propto - \bm k \cdot \bm p$ implies that low-momentum modes are essentially non-interacting, while high-momentum modes can scatter so strongly as to enter a collective hydrodynamic regime.
This separation between momentum sectors with distinct transport properties necessitates a kinetic framework that treats all momentum modes on equal footing.

Applied to atomically thin Heisenberg ferromagnets such as CrCl$_3$, our approach predicts a temperature-driven crossover accessible to qubit defects such as NV centers [Fig.~\ref{fig:schematic}(c)].
At low temperatures the magnon gas is dilute and only weakly interacting long-wavelength modes contribute, yielding a ballistic response whose noise grows with $T$. 
At high temperatures, strongly interacting short-wavelength modes become appreciably populated and dominate the probed signal, producing a hydrodynamic response whose noise decreases with $T$.
This anomaly arises because the NV center probes the tail of the sound resonance, whose magnetic signal is governed by a magnon viscosity that behaves as $\mu \sim 1/T$---a direct fingerprint of the SU(2)-symmetric exchange interaction~\cite{dyson_general_1956}.
This naturally accounts for the anomalous noise reported in CrCl$_3$~\cite{xue_magnon_2026} and predicts that cooling from $\sim$5\,K to $\sim$1\,K should map the complete evolution from the viscous hydrodynamic regime to ballistic quasiparticle dynamics.

These results are not specific to CrCl$_3$. We anticipate similar physics in other thin ferromagnets---CrBr$_3$~\cite{kim_evolution_2019,xing_magnon_2019}, ultrathin EuO and EuS~\cite{mauger_magnetic_1986,wei_strong_2016}, and Cr$_2$Ge$_2$Te$_6$~\cite{gong_discovery_2017}---and, moreover, the method is not confined to ferromagnets and applies to any 2D system describable by the QBE. 
The flexural phonons of suspended graphene, which combine a quadratic dispersion with a soft scattering phase space, offer a compelling non-magnetic analog~\cite{cepellotti_phonon_2015}, while the antiferromagnet MnPSe$_3$, with its linear Goldstone dispersion, would test whether the viscous anomaly persists beyond the ferromagnetic case~\cite{calder_magentic_2021}.

Several extensions stand out. The framework generalizes naturally to multi-mode systems, opening a landscape of richer transport regimes: interlayer coupling in multilayer CrCl$_3$ would clarify the role of heat conductivity in recent driven experiments~\cite{xue_magnon_2026}, while interacting Bose--Fermi mixtures---relevant to atomically thin semiconductors hosting both excitons and doped charges---would open the physics of mutual drag and entrainment between the two species, effects inaccessible to either alone.
The angular-harmonic basis likewise carries over to three dimensions through a spherical-harmonic expansion. 
Most ambitiously, the parabolic dispersion on which our Gaussian mixture representation rests can itself be relaxed: generalizing the collision-integral machinery to realistic, nonparabolic bands would bring complex materials within reach and, ultimately, enable inverse design of tailored response functions.

\begin{acknowledgments}
The authors would like to thank I.~Esterlis, H.~Le, R.~Xue, N.~Maksimovic, P.~Put, M.~D.~Lukin, A.~Yacoby, B.~Halperin, A.~G\'omez Salvador, J.~B.~Curtis, F.~Marijanovi\'c, and F.~Machado for fruitful discussions.

A.~M., O.~M. and E.~D. acknowledge support from the SNSF project 200021\_212899, the Swiss State Secretariat for Education, Research and Innovation (contract number UeM019-1), the SNSF Sinergia grant CRSII--222792, and NCCR SPIN, a National Centre of Competence in Research, funded by the Swiss National Science Foundation (grant number 225153). P.~E.~D. acknowledges support from the Quantum Optics Fellowship at UMD and the NSF QLCI (award No.~OMA-2120757).
\end{acknowledgments}

\bibliography{biblio}

@article{kim_evolution_2019,
  title={Evolution of interlayer and intralayer magnetism in three atomically thin chromium trihalides},
  author={Kim, Hyun Ho and Yang, Bowen and Li, Siwen and Jiang, Shengwei and Jin, Chenhao and Tao, Zui and Nichols, George and Sfigakis, Francois and Zhong, Shazhou and Li, Chenghe and others},
  journal={PNAS},
  volume={116},
  number={23},
  pages={11131--11136},
  year={2019},
  publisher={National Acad Sciences},
  doi = {10.1073/pnas.1902100116},
  url = {https://doi.org/10.1073/pnas.1902100116}
}

@article{rodriguez-nieva_probing_2022,
    title = {Probing hydrodynamic sound modes in magnon fluids using spin magnetometers},
    volume = {105},
    url = {https://link.aps.org/doi/10.1103/PhysRevB.105.174412},
    doi = {10.1103/PhysRevB.105.174412},
    number = {17},
    urldate = {2023-03-30},
    journal = {Physical Review B},
    publisher = {American Physical Society},
    author = {Rodriguez-Nieva, Joaquin F. and Podolsky, Daniel and Demler, Eugene},
    month = may,
    year = {2022},
    pages = {174412},
}

@article{massignan2005viscous,
  title={Viscous relaxation and collective oscillations in a trapped Fermi gas near the unitarity limit},
  author={Massignan, P and Bruun, Georg Morten and Smith, Henrik},
  journal={Phys. Rev. A},
  volume={71},
  number={3},
  pages={033607},
  year={2005},
  publisher={APS},
  url = {https://journals.aps.org/pra/pdf/10.1103/PhysRevA.71.033607},
  doi = {10.1103/PhysRevA.71.033607}
}

@article{dolgirev_accelerating_2024,
    title = {Accelerating analysis of {Boltzmann} equations using {Gaussian} mixture models: {Application} to quantum {Bose}-{Fermi} mixtures},
    volume = {6},
    shorttitle = {Accelerating analysis of {Boltzmann} equations using {Gaussian} mixture models},
    url = {https://link.aps.org/doi/10.1103/PhysRevResearch.6.033017},
    doi = {10.1103/PhysRevResearch.6.033017},
    number = {3},
    urldate = {2026-05-05},
    journal = {Physical Review Research},
    publisher = {American Physical Society},
    author = {Dolgirev, Pavel E. and Seetharam, Kushal and Kanász-Nagy, Márton and Robens, Carsten and Yan, Zoe Z. and Zwierlein, Martin and Demler, Eugene},
    month = jul,
    year = {2024},
    pages = {033017},
}

@article{halperin_hydrodynamic_1969,
    title = {Hydrodynamic {Theory} of {Spin} {Waves}},
    volume = {188},
    url = {https://link.aps.org/doi/10.1103/PhysRev.188.898},
    doi = {10.1103/PhysRev.188.898},
    number = {2},
    urldate = {2023-03-30},
    journal = {Physical Review},
    publisher = {American Physical Society},
    author = {Halperin, B. I. and Hohenberg, P. C.},
    month = dec,
    year = {1969},
    pages = {898--918},
}

@article{machado_quantum_2022,
  title = {Quantum Noise Spectroscopy of Dynamical Critical Phenomena},
  author = {Machado, Francisco and Demler, Eugene A. and Yao, Norman Y. and Chatterjee, Shubhayu},
  journal = {Phys. Rev. Lett.},
  volume = {131},
  issue = {7},
  pages = {070801},
  numpages = {8},
  year = {2023},
  month = {Aug},
  publisher = {American Physical Society},
  doi = {10.1103/PhysRevLett.131.070801},
  url = {https://link.aps.org/doi/10.1103/PhysRevLett.131.070801}
}

@article{iacocca_perspectives_2019,
	title = {Perspectives on spin hydrodynamics in ferromagnetic materials},
	volume = {383},
	issn = {03759601},
	url = {https://linkinghub.elsevier.com/retrieve/pii/S0375960119306619},
	doi = {10.1016/j.physleta.2019.125858},
	number = {28},
	urldate = {2023-08-03},
	journal = {Physics Letters A},
	author = {Iacocca, Ezio and Hoefer, Mark A.},
	month = oct,
	year = {2019},
	pages = {125858},
}

@article{bruno_spin_1991,
  title={Spin-wave theory of two-dimensional ferromagnets in the presence of dipolar interactions and magnetocrystalline anisotropy},
  author={Bruno, Patrick},
  journal={Phys. Rev. B},
  volume={43},
  number={7},
  pages={6015},
  year={1991},
  publisher={APS},
  doi = {10.1103/PhysRevB.43.6015},
  url = {https://doi.org/10.1103/PhysRevB.43.6015}
}

@Book{Mattis2006,
author={Mattis, Daniel C.},
title={Theory Of Magnetism Made Simple},
year={2006},
publisher={World Scientific Publishing Company},
address={Singapore},
url = {https://www.worldscientific.com/worldscibooks/10.1142/5372?srsltid=AfmBOorSTXMwf7KCDM403N-CV1hOxSX9065xdH9uyatcupdJpo_YZugV},
}

@book{chapman1990mathematical,
  author    = {Chapman, Sydney and Cowling, Thomas George},
  title     = {The Mathematical Theory of Non-Uniform Gases: An Account of the Kinetic Theory of Viscosity, Thermal Conduction and Diffusion in Gases},
  edition   = {3},
  series    = {Cambridge Mathematical Library},
  publisher = {Cambridge University Press},
  year      = {1990},
  isbn      = {978-0-521-40844-8},
  url       = {https://www.cambridge.org/9780521408448},
}

@book{kardar2007statistical,
    title = {Statistical {Physics} of {Particles}},
    url = {https://www.cambridge.org/highereducation/books/statistical-physics-of-particles/3CC2F33BD9F8DC56758DBDDB5B870558},
    urldate = {2026-07-10},
    journal = {Cambridge Aspire website},
    publisher = {Cambridge University Press},
    author = {Kardar, Mehran},
    month = jun,
    year = {2007},
    doi = {10.1017/CBO9780511815898}
}

@article{PhysRevLett.132.246504,
  title = {Local Noise Spectroscopy of Wigner Crystals in Two-Dimensional Materials},
  author = {Dolgirev, Pavel E. and Esterlis, Ilya and Zibrov, Alexander A. and Lukin, Mikhail D. and Giamarchi, Thierry and Demler, Eugene},
  journal = {Phys. Rev. Lett.},
  volume = {132},
  issue = {24},
  pages = {246504},
  numpages = {8},
  year = {2024},
  month = {Jun},
  publisher = {American Physical Society},
  doi = {10.1103/PhysRevLett.132.246504},
  url = {https://link.aps.org/doi/10.1103/PhysRevLett.132.246504}
}

@book{dorfman_contemporary_2021,
    address = {Cambridge},
    title = {Contemporary {Kinetic} {Theory} of {Matter}},
    isbn = {978-0-521-89547-7},
    url = {https://www.cambridge.org/core/books/contemporary-kinetic-theory-of-matter/393100F9C3F0678DD3884E2EF300D8B4},
    doi = {10.1017/9781139025942},
    urldate = {2026-07-10},
    publisher = {Cambridge University Press},
    author = {Dorfman, J. R. and van Beijeren, Henk and Kirkpatrick, T. R.},
    year = {2021},
}

@article{xue_magnon_2026,
    title = {Magnon hydrodynamics in an atomically thin ferromagnet},
    volume = {392},
    url = {https://www.science.org/doi/10.1126/science.adp2397},
    doi = {10.1126/science.adp2397},
    number = {6800},
    urldate = {2026-06-26},
    journal = {Science},
    publisher = {American Association for the Advancement of Science},
    author = {Xue, Ruolan and Maksimovic, Nikola and Dolgirev, Pavel E. and Xia, Li-Qiao and Müller, Aaron and Kitagawa, Ryota and Machado, Francisco and Klein, Dahlia R. and MacNeill, David and Watanabe, Kenji and Taniguchi, Takashi and Jarillo-Herrero, Pablo and Lukin, Mikhail D. and Demler, Eugene and Yacoby, Amir},
    month = may,
    year = {2026},
    pages = {873--878},
}

@article{Mermin1966,
  title = {Absence of Ferromagnetism or Antiferromagnetism in One- or Two-Dimensional Isotropic Heisenberg Models},
  author = {Mermin, N. D. and Wagner, H.},
  journal = {Phys. Rev. Lett.},
  volume = {17},
  issue = {22},
  pages = {1133--1136},
  numpages = {0},
  year = {1966},
  month = {Nov},
  publisher = {American Physical Society},
  doi = {10.1103/PhysRevLett.17.1133},
  url = {https://link.aps.org/doi/10.1103/PhysRevLett.17.1133}
}

@article{dyson_general_1956,
    title = {General {Theory} of {Spin}-{Wave} {Interactions}},
    volume = {102},
    url = {https://link.aps.org/doi/10.1103/PhysRev.102.1217},
    doi = {10.1103/PhysRev.102.1217},
    number = {5},
    urldate = {2023-08-17},
    journal = {Physical Review},
    author = {Dyson, Freeman J.},
    month = jun,
    year = {1956},
    pages = {1217--1230},
}

@article{dyson_thermodynamic_1956,
    title = {Thermodynamic {Behavior} of an {Ideal} {Ferromagnet}},
    volume = {102},
    url = {https://link.aps.org/doi/10.1103/PhysRev.102.1230},
    doi = {10.1103/PhysRev.102.1230},
    number = {5},
    urldate = {2023-08-17},
    journal = {Physical Review},
    author = {Dyson, Freeman J.},
    month = jun,
    year = {1956},
    pages = {1230--1244},
}

@article{cepellotti_phonon_2015,
	title = {Phonon hydrodynamics in two-dimensional materials},
	volume = {6},
	rights = {2015 Springer Nature Limited},
	issn = {2041-1723},
	url = {https://www.nature.com/articles/ncomms7400},
	doi = {10.1038/ncomms7400},
	pages = {6400},
	number = {1},
	journal = {Nature Communications},
	shortjournal = {Nat Commun},
	publisher = {Nature Publishing Group},
	author = {Cepellotti, Andrea and Fugallo, Giorgia and Paulatto, Lorenzo and Lazzeri, Michele and Mauri, Francesco and Marzari, Nicola},
	urldate = {2026-05-20},
	year = {2015},
	langid = {english},
}

@article{chumak_magnon_2015,
    title = {Magnon spintronics},
    volume = {11},
    copyright = {2014 Springer Nature Limited},
    issn = {1745-2481},
    url = {https://www.nature.com/articles/nphys3347},
    doi = {10.1038/nphys3347},
    number = {6},
    urldate = {2026-05-04},
    journal = {Nature Physics},
    publisher = {Nature Publishing Group},
    author = {Chumak, A. V. and Vasyuchka, V. I. and Serga, A. A. and Hillebrands, B.},
    month = jun,
    year = {2015},
    pages = {453--461},
}

@article{cornelissen_long-distance_2015,
	title = {Long-distance transport of magnon spin information in a magnetic insulator at room temperature},
	volume = {11},
	rights = {2015 Springer Nature Limited},
	issn = {1745-2481},
	url = {https://www.nature.com/articles/nphys3465},
	doi = {10.1038/nphys3465},
	pages = {1022--1026},
	number = {12},
	journal = {Nature Physics},
	shortjournal = {Nature Phys},
	publisher = {Nature Publishing Group},
	author = {Cornelissen, L. J. and Liu, J. and Duine, R. A. and Youssef, J. Ben and van Wees, B. J.},
	urldate = {2026-05-20},
	year = {2015},
	langid = {english},
}

@article{burch_magnetism_2018,
    title = {Magnetism in two-dimensional van der Waals materials},
    volume = {563},
    copyright = {2018 Springer Nature Limited},
    issn = {1476-4687},
    url = {https://www.nature.com/articles/s41586-018-0631-z},
    doi = {10.1038/s41586-018-0631-z},
    number = {7729},
    urldate = {2024-03-11},
    journal = {Nature},
    publisher = {Nature Publishing Group},
    author = {Burch, Kenneth S. and Mandrus, David and Park, Je-Geun},
    month = nov,
    year = {2018},
    pages = {47--52},
}

@article{cai_atomically_2019,
    title = {Atomically {Thin} {CrCl3}: {An} {In}-{Plane} {Layered} {Antiferromagnetic} {Insulator}},
    volume = {19},
    issn = {1530-6984},
    shorttitle = {Atomically {Thin} {CrCl3}},
    url = {https://doi.org/10.1021/acs.nanolett.9b01317},
    doi = {10.1021/acs.nanolett.9b01317},
    number = {6},
    urldate = {2025-10-01},
    journal = {Nano Letters},
    publisher = {American Chemical Society},
    author = {Cai, Xinghan and Song, Tiancheng and Wilson, Nathan P. and Clark, Genevieve and He, Minhao and Zhang, Xiaoou and Taniguchi, Takashi and Watanabe, Kenji and Yao, Wang and Xiao, Di and McGuire, Michael A. and Cobden, David H. and Xu, Xiaodong},
    month = jun,
    year = {2019},
    pages = {3993--3998},
}

@article{xing_magnon_2019,
    title = {Magnon {Transport} in {Quasi}-{Two}-{Dimensional} van der {Waals} {Antiferromagnets}},
    volume = {9},
    url = {https://link.aps.org/doi/10.1103/PhysRevX.9.011026},
    doi = {10.1103/PhysRevX.9.011026},
    number = {1},
    urldate = {2026-01-26},
    journal = {Physical Review X},
    publisher = {American Physical Society},
    author = {Xing, Wenyu and Qiu, Luyi and Wang, Xirui and Yao, Yunyan and Ma, Yang and Cai, Ranran and Jia, Shuang and Xie, X.C. and Han, Wei},
    month = feb,
    year = {2019},
    pages = {011026},
}

@article{serga_yig_2010,
    title = {{YIG} magnonics},
    volume = {43},
    issn = {0022-3727},
    url = {https://dx.doi.org/10.1088/0022-3727/43/26/264002},
    doi = {10.1088/0022-3727/43/26/264002},
    number = {26},
    urldate = {2023-07-30},
    journal = {Journal of Physics D: Applied Physics},
    author = {Serga, A. A. and Chumak, A. V. and Hillebrands, B.},
    month = jun,
    year = {2010},
    pages = {264002},
}

@article{mauger_magnetic_1986,
    title = {The magnetic, optical, and transport properties of representatives of a class of magnetic semiconductors: {The} europium chalcogenides},
    volume = {141},
    issn = {0370-1573},
    shorttitle = {The magnetic, optical, and transport properties of representatives of a class of magnetic semiconductors},
    url = {https://www.sciencedirect.com/science/article/pii/0370157386901390},
    doi = {10.1016/0370-1573(86)90139-0},
    number = {2},
    urldate = {2026-06-26},
    journal = {Physics Reports},
    author = {Mauger, A. and Godart, C.},
    month = aug,
    year = {1986},
    pages = {51--176},
}

@article{wei_strong_2016,
    title = {Strong interfacial exchange field in the graphene/{EuS} heterostructure},
    volume = {15},
    issn = {1476-4660},
    doi = {10.1038/nmat4603},
    number = {7},
    journal = {Nature Materials},
    author = {Wei, Peng and Lee, Sunwoo and Lemaitre, Florian and Pinel, Lucas and Cutaia, Davide and Cha, Wujoon and Katmis, Ferhat and Zhu, Yu and Heiman, Donald and Hone, James and Moodera, Jagadeesh S. and Chen, Ching-Tzu},
    month = jul,
    year = {2016},
    pages = {711--716},
}

@article{calder_magentic_2021,
  title = {Magnetic exchange interactions in the van der Waals layered antiferromagnet $\mathrm{Mn}\mathrm{P}{\mathrm{Se}}_{3}$},
  author = {Calder, S. and Haglund, A. V. and Kolesnikov, A. I. and Mandrus, D.},
  journal = {Phys. Rev. B},
  volume = {103},
  issue = {2},
  pages = {024414},
  numpages = {9},
  year = {2021},
  month = {Jan},
  publisher = {American Physical Society},
  doi = {10.1103/PhysRevB.103.024414},
  url = {https://link.aps.org/doi/10.1103/PhysRevB.103.024414}
}

@article{gurzhi_1968,
	author = {R. N. Gurzhi},
	title = {Hydrodynamic effects in solids at low temperature},
	publisher = {Physics-Uspekhi},
	year = {1968},
	journal = {Phys. Usp.},
	volume = {11},
	number = {2},
	pages = {255-270},
	url = {https://ufn.ru/en/articles/1968/2/g/},
	doi = {10.1070/PU1968v011n02ABEH003815}
}

@article{Basso2016,
  title = {Thermodynamic transport theory of spin waves in ferromagnetic insulators},
  author = {Basso, Vittorio and Ferraro, Elena and Piazzi, Marco},
  journal = {Phys. Rev. B},
  volume = {94},
  pages = {144422},
  year = {2016},
  doi = {10.1103/PhysRevB.94.144422}
}

@article{CepellottiMarzari2016,
  title = {Thermal Transport in Crystals as a Kinetic Theory of Relaxons},
  author = {Cepellotti, Andrea and Marzari, Nicola},
  journal = {Phys. Rev. X},
  volume = {6},
  pages = {041013},
  year = {2016},
  doi = {10.1103/PhysRevX.6.041013}
}

@article{PhysRevB.88.045430,
  title = {Ab initio variational approach for evaluating lattice thermal conductivity},
  author = {Fugallo, Giorgia and Lazzeri, Michele and Paulatto, Lorenzo and Mauri, Francesco},
  journal = {Phys. Rev. B},
  volume = {88},
  issue = {4},
  pages = {045430},
  numpages = {9},
  year = {2013},
  month = {Jul},
  publisher = {American Physical Society},
  doi = {10.1103/PhysRevB.88.045430},
  url = {https://link.aps.org/doi/10.1103/PhysRevB.88.045430}
}

@article{du2017control,
  title = {Control and local measurement of the spin chemical potential in a magnetic insulator},
  author = {Du, Chunhui and van der Sar, Toeno and Zhou, Tony X. and Upadhyaya, Pramey and Casola, Francesco and Zhang, Huiliang and Onbasli, Mehmet C. and Ross, Caroline A. and Walsworth, Ronald L. and Tserkovnyak, Yaroslav and Yacoby, Amir},
  journal = {Science},
  volume = {357},
  pages = {195},
  year = {2017},
  doi = {10.1126/science.aak9611}
}

@article{wang_noninvasive_2022,
    title = {Noninvasive measurements of spin transport properties of an antiferromagnetic insulator},
    volume = {8},
    url = {https://www.science.org/doi/10.1126/sciadv.abg8562},
    doi = {10.1126/sciadv.abg8562},
    number = {1},
    urldate = {2026-08-14},
    journal = {Science Advances},
    publisher = {American Association for the Advancement of Science},
    author = {Wang, Hailong and Zhang, Shu and McLaughlin, Nathan J. and Flebus, Benedetta and Huang, Mengqi and Xiao, Yuxuan and Liu, Chuanpu and Wu, Mingzhong and Fullerton, Eric E. and Tserkovnyak, Yaroslav and Du, Chunhui Rita},
    month = jan,
    year = {2022},
    pages = {eabg8562},
}

@article{broido_intrinsic_2007,
    title = {Intrinsic lattice thermal conductivity of semiconductors from first principles},
    volume = {91},
    issn = {0003-6951},
    url = {https://doi.org/10.1063/1.2822891},
    doi = {10.1063/1.2822891},
    number = {23},
    urldate = {2026-08-14},
    journal = {Applied Physics Letters},
    author = {Broido, D. A. and Malorny, M. and Birner, G. and Mingo, Natalio and Stewart, D. A.},
    month = dec,
    year = {2007},
    pages = {231922},
}

@article{gong_discovery_2017,
    title = {Discovery of intrinsic ferromagnetism in two-dimensional van der {Waals} crystals},
    volume = {546},
    copyright = {2017 Macmillan Publishers Limited, part of Springer Nature. All rights reserved.},
    issn = {1476-4687},
    url = {https://www.nature.com/articles/nature22060},
    doi = {10.1038/nature22060},
    number = {7657},
    urldate = {2026-08-14},
    journal = {Nature},
    publisher = {Nature Publishing Group},
    author = {Gong, Cheng and Li, Lin and Li, Zhenglu and Ji, Huiwen and Stern, Alex and Xia, Yang and Cao, Ting and Bao, Wei and Wang, Chenzhe and Wang, Yuan and Qiu, Z. Q. and Cava, R. J. and Louie, Steven G. and Xia, Jing and Zhang, Xiang},
    month = jun,
    year = {2017},
    pages = {265--269},
}

\clearpage
\newpage

\onecolumngrid
\appendix

\section{  Derivation of collision-integral matrix elements} \label{sup:matrix_elements}

In this section, we explicitly evaluate the magnon decay rate $\gamma$ and the collision kernels ${\cal M}^{\rm II}$ and ${\cal M}^{\rm III}$ entering the linearized QBE~\eqref{eq:LBE_angular_harmonics}, using the Gaussian Mixture Model (GMM) representation~\cite{dolgirev_accelerating_2024} of the equilibrium distribution function $n_p^{(0)}$ in Eq.~\eqref{eq:GMM}. 
We also summarize the numerical implementation in the \texttt{LBE-GMM-solver} repository.

\subsection{Analysis of the magnon decay rate $\gamma$} \label{sec:decay_rate}

The expressions for the magnon decay rates corresponding to the contact [Eq.~\eqref{eq:Tmatrix_contact}] and exchange [Eq.~\eqref{eq:Tmatrix_exchange}] interactions follow directly from Eq.~\eqref{eqn:CI_gen_exc}:
\begin{align}
    \tilde \gamma_{\rm c}(p)  =& \frac{\tilde{g}_{\rm c}^2}{ \Delta_p}\int \frac{d^2 \bm k}{(2\pi)^2}\int\frac{d^2 \bm p'}{(2\pi)^2} \int\frac{d^2 \bm k'}{(2\pi)^2} (2\pi)\delta( \varepsilon_{p} + \varepsilon_{k} - \varepsilon_{p'} - \varepsilon_{k'})  (2\pi)^2\delta(\bm p + \bm k - {\bm p'} - {\bm k'}) \notag \\
    &\qquad \times n^{(0)}_{ p} n^{(0)}_{ k } (1 + n^{(0)}_{ p'}) (1 + n^{(0)}_{ k'}), \\
    \tilde \gamma_{\rm ex}(p) =&  \frac{\tilde{g}_{\rm ex}^2}{\Delta_p}\int \frac{d^2 \bm k}{(2\pi)^2}\int\frac{d^2 \bm p'}{(2\pi)^2} \int\frac{d^2 \bm k'}{(2\pi)^2}  |\bm k\cdot \bm p|^2  (2\pi) \delta(\varepsilon_{p} + \varepsilon_{k} - \varepsilon_{p'} - \varepsilon_{k'}) (2\pi)^2\delta\left(\bm p + \bm k - \bm p' - \bm k'\right) \notag \\
    &\qquad \times n^{(0)}_{ p} n^{(0)}_{ k } (1 + n^{(0)}_{ p'}) (1 + n^{(0)}_{ k'}).
\end{align}
Transforming to the center-of-mass (CoM) frame $\bm p = \bm P/2 + \bm q$, $\bm k = \bm P/2 - \bm q$, $\bm p' = \bm P/2 + \bm q'$, $\bm k' = \bm P/2 - \bm q'$, we obtain:
\begin{align}
   \gamma_{\rm int}(p) &=  \frac{\tilde{g}_{\rm int}^2}{n^{(0)}_p}\int \frac{d^2 \bm P}{(2\pi)^2}\int\frac{d^2 \bm q}{(2\pi)^2} \int\frac{d^2 \bm q'}{(2\pi)^2} {\text{Poly}}_{\rm int}[\bm P,\bm q,\bm q']  (2\pi) \delta( q^2 - q'^2 )  (2\pi)^2\delta(\bm p - \bm P/2 - \bm q) \, (1+n^{(0)}_{ k }) n^{(0)}_{k'} n^{(0)}_{ p'} . \label{eq:gamma_p}
\end{align}
Here, the polynomials ${\text{Poly}}_{\rm ex} = (P^2/4 -q^2)^2$ and ${\text{Poly}}_{\rm c} = 1$ encode the scattering vertex structure.

Using the GMM expansion~\eqref{eq:GMM}, the evaluation of Eq.~\eqref{eq:gamma_p} reduces to integrals of the form:
\begin{align}
    \int \frac{d^2 \bm P}{(2\pi)^2}\int\frac{d^2 \bm q}{(2\pi)^2} \int\frac{d^2 \bm q'}{(2\pi)^2}   {\text{Poly}}[\bm P,\bm q,\bm q']
      (2\pi) & \delta( q^2 - q'^2 ) (2\pi)^2\delta(\bm p - \bm P/2 - \bm q) \notag\\
      & \times \exp \Big\{ - \frac{P^2}{2\gamma_{P}^2} - \frac{q^2}{\gamma_q^2} + \frac{1}{\gamma_P} (A \bm q + B \bm q') \cdot \bm P \Big\}.
\end{align}
Integrating out the CoM momentum $\bm P$ yields:
\begin{align}
    4 \exp\Big\{ -\frac{2p^2}{\gamma_P^2} \Big\} \int\frac{d^2 \bm q}{(2\pi)^2} \int\frac{d^2 \bm q'}{(2\pi)^2}   {\text{Poly}}[\bm q,\bm q'] (2\pi) & \delta( q^2 - q'^2 ) \exp \{ - C_1 q^2 + C_2 \bm q \cdot\bm q' + (\alpha \bm q + \beta \bm q')\cdot\bm p \} , \label{eq:gamma_p_int}
\end{align}
with coefficients $C_1 = 2/\gamma_P^2 + 1/\gamma_q^2 + 2A/\gamma_P$, $C_2 = - 2 B/\gamma_P$, $\alpha = 2A/\gamma_P + 4/\gamma_P^2$,  $\beta = 2B/\gamma_P$. To disentangle the scalar product $\bm q\cdot \bm q'$, we perform the linear transformation derived in Ref.~\cite{dolgirev_accelerating_2024}:
\begin{align}
    C_1 q^2 - C_2 \bm q\cdot \bm q' = (\tilde{A}\bm q + \tilde{B} \bm q')^2,
\end{align}
defined by the coefficients:
\begin{align}
    \Tilde{A} = \sqrt{ \frac{1}{2} \Big( C_1  + \sqrt{C_1^2 - C_2^2} \Big)},\qquad \Tilde{B} = - \text{sign}(C_2) \sqrt{ \frac{1}{2} \Big( C_1  - \sqrt{C_1^2 - C_2^2} \Big)} . 
\end{align}
In terms of the new variables $\bm k = \tilde{A}\bm q + \tilde{B} \bm q'$ and $\bm k' = \tilde{A}\bm q' + \tilde{B} \bm q$, we get:
\begin{align}
    \delta(q^2-q'^2) = (\tilde{A}^2 - \tilde{B}^2)\delta(k^2-k'^2),\qquad \int\frac{d^2\bm q}{(2\pi)^2}\int\frac{d^2\bm q'}{(2\pi)^2} \to \frac{1}{(\tilde{A}^2 - \tilde{B}^2)^2} \int\frac{d^2\bm k}{(2\pi)^2}\int\frac{d^2\bm k'}{(2\pi)^2}.
\end{align}
Equation~\eqref{eq:gamma_p_int} then simplifies to:
\begin{align}
    4 \frac{\exp\{ -2p^2/\gamma_P^2\} }{\tilde{A}^2 - \tilde{B}^2}  \int\frac{d^2\bm k}{(2\pi)^2}\int\frac{d^2\bm k'}{(2\pi)^2} \text{Poly}[\bm k,\bm k'] (2\pi) \delta(k^2 - k'^2) \exp\{-k^2 + (\tilde{\alpha}\bm k + \tilde{\beta}\bm k')\cdot\bm p \},
\end{align}
where 
\begin{align}
    \tilde \alpha = \frac{\alpha \tilde A-\beta \tilde B}{\tilde A^2-\tilde B^2} \, ,\qquad
    \tilde \beta = \frac{\beta \tilde A-\alpha \tilde B}{\tilde A^2-\tilde B^2}. 
\end{align}
The angular integrals can be performed analytically. For the contact interaction, this yields:
\begin{align}
    2 \frac{\exp\{ -2p^2/\gamma_P^2\} }{\tilde{A}^2 - \tilde{B}^2}  \int_0^\infty\frac{d k\, k}{2\pi} \exp\{-k^2 \} I_0(\tilde{\alpha} k p) I_0( \tilde{\beta}kp) .
\end{align}
For the exchange interaction, the corresponding angular integral is evaluated symbolically using a short {\it Mathematica} routine. The remaining $k$-integrals are then computed numerically.
For reference, the classical decay rates are $\gamma_{\rm c}(p) \approx z \tilde{g}_c^2/(4\pi)$ for the contact interaction and $\gamma_{\rm ex}^{\rm cl}(p) = z \tilde{g}_{\rm ex}^2 p^2/4\pi$ for the exchange interaction; these expressions serve as useful benchmarks for our numerical results.

Figure~\ref{fig:schematic}(b) shows the resulting magnon scattering rates. The contact decay rate remains finite as $\tilde{p} \to 0$, whereas the exchange decay rate vanishes, reflecting the kinematic suppression of the soft scattering vertex.

\subsection{The ballistic-to-hydrodynamic crossover estimates}
\label{sec:crossover_estimate}

To estimate the crossover temperature, we introduce the dimensionless ratio:
\begin{align}
    {\cal R}_T(T;q,\omega)=\frac{\gamma_{\rm ex}(p_T,T)}{\sqrt{q^2T/m+\omega^2}},
    \label{eq:hydrodynamicity_ratio}
\end{align}
which satisfies ${\cal R}_T\gg1$ deep in the hydrodynamic regime.
Restoring dimensions in the exchange decay rate computed above gives
\begin{align}
    \gamma_{\rm ex}(p_T,T)=m^4T^3g_{\rm ex}^2{\cal G}(1,z),
\end{align}
where ${\cal G}$ is the dimensionless collision integral evaluated at $\tilde p=1$. Its GMM evaluation is accurately approximated by
\begin{align}
    {\cal G}(1,z)\simeq c_{\cal G}g_1(z),    \qquad c_{\cal G}\simeq0.078,    \qquad g_1(z)=-\ln(1-z).
\end{align}
For $\Delta\ll T$, we further have $g_1(e^{-\Delta/T})\simeq|\ln(\Delta/T)|$. 

The crossover condition is obtained as ${\cal R}_T(T_\times;q_{\rm NV},\omega_{\rm NV})={\cal R}_\times$, with the probe filter functions selecting $q_{\rm NV}=3/(2d_{\rm NV})$ and $\omega_{\rm NV}\simeq4.6/\tau$.
Using $|g_{\rm ex}|=3\sqrt{2}{\cal A}_{\rm uc}Ja^2/4$, calibration against the full QBE crossover for $\Delta=0.17\,$K and $0.32\,$K fixes ${\cal R}_\times\simeq0.29$.

\subsection{Analysis of ${\cal M}^{\rm II}$}
As follows from Eq.~\eqref{eqn:C_II_gen_exc}, the kernel ${\cal M}^{\rm II}$ takes the form:
\begin{gather} 
    {\cal M}^{\rm II}_{{\rm c}, p p'}(\vartheta)  =  -\tilde{g}_c^2 M^{\rm II}_{ p p'}(\vartheta) ,\label{eq:MatII_contact}  \\
    {\cal M}^{\rm II}_{{\rm ex}, p p'}(\vartheta)  =-  \tilde{g}_{\rm ex}^2  p^2 p'^2 \cos^2(\vartheta) M^{\rm II}_{ p p'}(\vartheta) \label{eq:MatII_exchange}.
\end{gather}
Using the GMM expansion~\eqref{eq:GMM}, we obtain for $M^{\rm II}$:
\begin{align}
    M^{\rm II}_{ p p'}(\vartheta) &= \frac{1 + n^{(0)}_{p'}}{n^{(0)}_p} \int \frac{d^2 \bm k}{(2\pi)^2}\int\frac{d^2 \bm k'}{(2\pi)^2} (2\pi) \delta( \varepsilon_{p} + \varepsilon_{p'} - \varepsilon_{k} - \varepsilon_{k'})  (2\pi)^2\delta(\bm p + \bm p' - \bm k - \bm k') \,  n^{(0)}_{ k} n^{(0)}_{ k'} \notag \\
    &= \frac{(1 + n^{(0)}_{p'})}{2 n^{(0)}_p} \sum_{s_1 s_2} a_{s_1}a_{s_2} \exp\Big\{ - \frac{(\gamma_{s_1}^2 + \gamma_{s_2}^2) (p^2 + p'^2)}{4\gamma_{s_1}^2\gamma_{s_2}^2}\Big\} I_0\Big\{ \frac{(\gamma_{s_1}^2-\gamma_{s_2}^2)}{4\gamma_{s_1}^2\gamma_{s_2}^2} \sqrt{(p^2 + p'^2)^2 - 4 p^2 p'^2 \cos^2(\vartheta)} \Big\},
\end{align}
where $\vartheta$ is the angle between $\bm p$ and $\bm p'$. In practice, we evaluate the angular Fourier harmonics of ${\cal M}^{\rm II}_{pp'}(\vartheta)$ numerically yielding ${\cal M}^{{\rm II},n}_{pp'}$.For reference, the classical limits are:
\begin{gather}
    {\cal M}_{{\rm c}, pp'}^{\rm II, n} \approx - \frac{1}{2}\delta_{n,0} \tilde{g}_c^2 z e^{-{p'}^2/2} ,\\
    {\cal M}_{{\rm ex},pp'}^{{\rm II},n} \approx - \frac{1}{4}\Big(\delta_{n,0} + \frac{\delta_{n,2}}{2}  + \frac{\delta_{n,-2}}{2}\Big) z\tilde{g}_{\rm ex}^2  p^2 p'^2 e^{-{p'}^2/2}. \label{eq:MII_classical}
\end{gather}
The integral action of this kernel is then:
\begin{align}
    {\cal K}^{{\rm II}}_{p}(\varphi) \coloneqq \int\frac{d^2\bm p'}{(2\pi)^2} {\cal M}^{\rm II}_{\bm p,\bm p'} \Phi_{\bm p'} =  \sum_m \int_0^\infty \frac{dp' p'}{2\pi} e^{i m \varphi} {\cal M}^{{\rm II},m}_{pp'}\,\Phi_{p'}^{m} \Rightarrow [{\cal O}^{\rm II}]^{n n'}_{p p'} =   \frac{p'}{2\pi} {\cal M}^{\rm II,n}_{p p'} \delta_{n,n'}. \label{eq:KII_phi} 
\end{align}

\subsection{Analysis of ${\cal M}^{\rm III}$}

\subsubsection{Contact Interaction}
As follows from Eq.~\eqref{eqn:C_II_gen_exc}, we can write:
\begin{align}
    {\cal M}^{\rm III}_{{\rm c}, \bm p\bm p'} & = 2\tilde{g}_c^2 \frac{ 1 + n^{(0)}_{p'}}{1 + n^{(0)}_{p}}\int \frac{d^2 \bm k}{(2\pi)^2}\int\frac{d^2 \bm k'}{(2\pi)^2}  (2\pi) \delta( \varepsilon_{p} + \varepsilon_k - \varepsilon_{p'} - \varepsilon_{k'}) (2\pi)^2\delta(\bm p + \bm k - \bm p' - \bm k') \,  n^{(0)}_{ k } (1 + n^{(0)}_{ k'})  \notag \\
    & = 2\tilde{g}_c^2 \frac{ 1 + n^{(0)}_{p'}}{1 + n^{(0)}_{p}} \int \frac{d^2 \bm P}{(2\pi)^2}  (2\pi) \delta( \bm P \cdot (\bm p - \bm p')) \,  n^{(0)}_{\bm P + \bm p'} (1 + n^{(0)}_{\bm P + \bm p}).
\end{align}
Decomposing $\bm P = \alpha \hat{n}_\parallel + \beta \hat{n}_{\perp}$ (with $\hat{n}_\parallel \parallel \bm p - \bm p'$), the $\alpha$-integral is trivialized by the delta function, yielding:
\begin{align}
    {\cal M}^{\rm III}_{{\rm c}, \bm p\bm p'} 
 & = \frac{2\tilde{g}_c^2}{|\bm p - \bm p'|} \frac{ 1 + n^{(0)}_{p'}}{1 + n^{(0)}_{p}} \int \frac{d \beta}{2\pi}   n^{(0)}_{\bm P + \bm p'} (1 + n^{(0)}_{\bm P + \bm p}). \label{eq:MatIII_contact_final}
\end{align}
The first term proportional to $n^{(0)}_{\bm P + \bm p'}$ can be easily computed 
using the GMM expansion~\equ{eq:GMM}:
\begin{align}
    \int \frac{d \beta}{2\pi}   n^{(0)}_{\bm P + \bm p'}  & = \sum_{s} a_s \int \frac{d \beta}{2\pi}  \exp\Big\{ -\frac{1}{2\gamma_s^2}(\beta \hat{n}_\perp + \bm p')^2 \Big\}   =
    \sum_{s} \frac{a_s \gamma_s}{\sqrt{2\pi}}  \exp \{ - p_\parallel'^2/(2\gamma_s^2) \},
\end{align}
where $\vartheta$ is the angle between $\bm p$ and $\bm p'$ and 
\begin{align}
    p_\parallel = \frac{p^2 - pp'\cos\vartheta}{|\bm p - \bm p'|},\qquad  p_\parallel' = \frac{ pp'\cos\vartheta - p'^2}{|\bm p - \bm p'|}, \qquad p_\perp = p_\perp' = \frac{ pp'\sin\vartheta}{|\bm p - \bm p'|}. \label{eq:para_and_perp_momentum_def}
\end{align}
The second term proportional to $n^{(0)}_{\bm P + \bm p'} n^{(0)}_{\bm P + \bm p}$ is found to be
\begin{align}
    \int \frac{d \beta}{2\pi}  n^{(0)}_{\bm P + \bm p'} n^{(0)}_{\bm P + \bm p}  =
    \sum_{s_1 s_2} \frac{a_{s_1} a_{s_2}\tilde{\gamma}}{\sqrt{2\pi}} \exp \{ - p_\parallel'^2/(2\gamma_{s_1}^2) - p_\parallel^2/(2\gamma_{s_2}^2) \},
\end{align}
where $\tilde{\gamma} = \gamma_{s_1}\gamma_{s_2}/\sqrt{\gamma_{s_1}^2 + \gamma_{s_2}^2}$. The classical limit reads:
\begin{align}
     {\cal M}^{\rm III, {\rm cl}}_{{\rm c}, \bm p\bm p'} 
 & \approx \frac{2\tilde{g}_c^2 }{|\bm p - \bm p'|}\frac{z}{\sqrt{2\pi}} \exp \{ - p_\parallel'^2/2 \}. \label{eq:MIII_contact_classical}
\end{align}

\subsubsection{Exchange Interaction}

For the exchange interaction, we obtain:
\begin{align}
    {\cal M}^{\rm III}_{{\rm ex},\bm p\bm p'} & = 2\tilde{g}_{\rm ex}^2  \frac{1 + n^{(0)}_{p'}}{1 + n^{(0)}_{p}}\int \frac{d^2 \bm k}{(2\pi)^2}\int\frac{d^2 \bm k'}{(2\pi)^2}  
    (\bm k\cdot \bm p)^2  (2\pi) \delta( \varepsilon_{p} + \varepsilon_k - \varepsilon_{p'} - \varepsilon_{k'}) (2\pi)^2\delta(\bm p + \bm k - \bm p' - \bm k') \,  n^{(0)}_{ k } (1 + n^{(0)}_{ k'}) \notag\\
    & = 2\tilde{g}_{\rm ex}^2  \frac{1 + n^{(0)}_{p'}}{1 + n^{(0)}_{p}}\int \frac{d^2 \bm P}{(2\pi)^2} ( \bm p\cdot (\bm P + \bm p'))^2 \, (2\pi) \delta( \bm P \cdot (\bm p - \bm p')) \,  n^{(0)}_{\bm P + \bm p'} (1 + n^{(0)}_{\bm P + \bm p}), 
\end{align}
Using the same coordinate transformation as in the contact case, one finds:
\begin{align}
    {\cal M}^{\rm III}_{{\rm ex},\bm p\bm p'} 
 & = \frac{2\tilde{g}_{\rm ex}^2 }{|\bm p - \bm p'|} \frac{(1 + n^{(0)}_{p'})}{(1 + n^{(0)}_{p})}\int \frac{d \beta}{2\pi} ( \bm p\cdot (\beta\hat{n}_\perp + \bm p'))^2 \,  n^{(0)}_{\bm P + \bm p'} (1 + n^{(0)}_{\bm P + \bm p}). \label{eq:MatIII_exchange_final}
\end{align}
Evaluating the $\beta$-integrals using the GMM expansion yields for the first term proportional to $n^{(0)}_{\bm P + \bm p'}$:
\begin{align}
    \int \frac{d \beta}{2\pi} ( \bm p\cdot (\beta\hat{n}_\perp + \bm p'))^2 \,  n^{(0)}_{\bm P + \bm p'}  & = \sum_{s} a_s \int \frac{d \beta}{2\pi} ( \bm p\cdot (\beta\hat{n}_\perp + \bm p'))^2 \, \exp\Big\{ -\frac{1}{2\gamma_s^2}(\beta \hat{n}_\perp + \bm p')^2 \Big\} \notag\\
    & =
    \sum_{s} \frac{a_s}{\sqrt{2\pi}} ( \gamma_s p_\parallel^2 p_\parallel'^2 +  \gamma_s^3 p_\perp^2 ) \exp \{ - p_\parallel'^2/(2\gamma_s^2) \}.
\end{align}
The second term proportional to $n^{(0)}_{\bm P + \bm p'} n^{(0)}_{\bm P + \bm p}$ reads:
\begin{align}
    \int \frac{d \beta}{2\pi} ( \bm p\cdot (\beta\hat{n}_\perp + \bm p'))^2 \,  n^{(0)}_{\bm P + \bm p'} n^{(0)}_{\bm P + \bm p} &= 
    \sum_{s_1 s_2}a_{s_1} a_{s_2} \int \frac{d \beta}{2\pi} ( \bm p\cdot (\beta\hat{n}_\perp + \bm p'))^2 \notag\\
    &  \times \exp\Big\{ -\frac{1}{2\gamma_{s_1}^2}(\beta \hat{n}_\perp + \bm p')^2  -\frac{1}{2\gamma_{s_2}^2}(\beta \hat{n}_\perp + \bm p)^2  \Big\} \notag\\
    & =
    \sum_{s_1 s_2} \frac{a_{s_1} a_{s_2}}{\sqrt{2\pi}} ( \tilde{\gamma} p_\parallel^2 p_\parallel'^2 +  \tilde{\gamma}^3 p_\perp^2 ) \exp \{ - p_\parallel'^2/(2\gamma_{s_1}^2) - p_\parallel^2/(2\gamma_{s_2}^2) \},
\end{align}
where $\tilde{\gamma} = \gamma_{s_1}\gamma_{s_2}/\sqrt{\gamma_{s_1}^2 + \gamma_{s_2}^2}$ and we used that $p_\perp = p_\perp'$. In the classical limit, we have:
\begin{align}
    {\cal M}^{\rm III, {\rm cl} }_{{\rm ex}, \bm p\bm p'} 
    & \approx \frac{2\tilde{g}_{\rm ex}^2 }{|\bm p - \bm p'|}\frac{z}{\sqrt{2\pi}} ( p_\parallel^2 p_\parallel'^2 +  p_\perp^2 ) \exp \{ - p_\parallel'^2/2 \}. \label{eq:MIII_exchange_classical}
\end{align}

\subsubsection{Numerical Considerations}

The kernel exhibits an integrable singularity ${\cal M}^{\rm III}_{\bm p\bm p'}\sim |\bm p - \bm p'|^{-1}$ for $|\bm p - \bm p'|\to 0$. To facilitate numerical integration, we therefore introduce the shifted variable $\bm p'' = \bm p' - \bm p$, which yields
\begin{align}
    {\cal K}^{\rm III}_{\bm p} \coloneqq \int\frac{d^2\bm p'}{(2\pi)^2} {\cal M}^{\rm III}_{\bm p,\bm p'} \Phi_{\bm p'} = \int\frac{d^2\bm p'}{(2\pi)^2} \frac{1}{|\bm p - \bm p'|} {\cal \bar{M}}^{\rm III}_{\bm p,\bm p'} \Phi_{\bm p'} =  \int_0^\infty\frac{d p''}{2\pi} \int_0^{2\pi}\frac{d \varphi''}{2\pi}  {\cal \bar{M}}^{\rm III}_{\bm p,\bm p + \bm p''} \Phi_{\bm p + \bm p''}. \label{eq:K_III_v0} 
\end{align}
In the new variables, we further have
\begin{gather}
    p_\parallel \to -p \cos(\varphi''),\qquad p_\parallel'\to - p'' -p \cos(\varphi''),\qquad p_\perp = p_\perp' \to p \sin(\varphi'').
\end{gather}
Although this transformation removes the divergence, it leaves the shifted deviation $\Phi_{\bm p+\bm p''}$, whose argument depends parametrically on both $(p,\varphi)$ and $(p'',\varphi'')$. In our numerical scheme, we construct this quantity by linear interpolation, writing $\Phi_{\bm k} \equiv \Phi_{\bm p+\bm p''}\equiv \Phi_{p,p''}(\varphi,\varphi'')$, where
\begin{align}
    k_x = p \cos(\varphi) + p'' \cos(\varphi''), \qquad
    k_y = p \sin(\varphi) + p'' \sin(\varphi'').
\end{align}
To evaluate the angular integral in Eq.~\eqref{eq:K_III_v0}, we again expand in Fourier harmonics:
\begin{align}
    {\cal \bar{M}}^{\rm III}_{\bm p,\bm p + \bm p''} = {\cal \bar{M}}^{\rm III}_{p, p''}(\varphi-\varphi'') \approx \sum_m e^{i m(\varphi-\varphi'')} {\cal \bar{M}}^{{\rm III},m}_{p, p''} , \qquad \Phi_{\bm p + \bm p''} = \Phi_{p,p''}(\varphi, \varphi'') \approx \sum_m e^{i m\varphi''} \Phi^{m}_{p,p''}(\varphi),
\end{align}
which yields
\begin{align}
    {\cal K}^{\rm III}_{p}(\varphi) = \sum_m \int_0^\infty\frac{d p''}{2\pi} e^{i m \varphi}  {\cal \bar{M}}^{{\rm III},m}_{p, p''}\Phi^{m}_{p,p''}(\varphi) .\label{eq:KIII_phi}
\end{align}
This is similar to the expression obtained for ${\cal K}^{\rm II}_p(\varphi)$ in Eq.~\eqref{eq:KII_phi}, except that here $\Phi$ retains an explicit $\varphi$ dependence because of the shifted argument $\bm p+\bm p''$. We further encode the linear interpolation in a matrix $U$ and write
\begin{align}
    {\cal K}^{\rm III}_{p}(\varphi) = \sum_{n'} \int_0^\infty\frac{d p'}{2\pi} \sum_m \int_0^\infty\frac{d p''}{2\pi} e^{i m \varphi}  {\cal \bar{M}}^{{\rm III},m}_{p, p''} U^{m, n'}_{p,p'',p'}(\varphi) \Phi^{n'}_{p'}.
\end{align}
This allows us to identify
\begin{align}
    \left[{\cal O}^{{\rm III}}\right]^{nn'}_{pp'}= \frac{1}{2 \pi} \sum_m \int_0^\infty\frac{d p''}{2\pi}  {\cal \bar{M}}^{{\rm III},m}_{p, p''} U^{n,m, n'}_{p,p'',p'} , \qquad  U^{n, m, n'}_{p,p'',p'} = \int \frac{d \varphi}{2 \pi} e^{i (m-n) \varphi} U^{m, n'}_{p,p'',p'}(\varphi).
\end{align}

\section{Variational Estimate of Transport Coefficients} \label{sec:microscopic_transport_coeff}

For completeness, we provide the explicit expressions for the method-of-moments variational lower bounds on the magnon transport coefficients~\cite{chapman1990mathematical,massignan2005viscous}; see Ref.~\cite{xue_magnon_2026} for a derivation:
\begin{gather}
    \mu \geq \frac{1 }{ T m^2} \Big[\int \frac{d^2 \bm p}{(2\pi)^2 } n^{(0)}_{\bm p} (1 + n^{(0)}_{\bm p})  p_x^2 p_y^2\Big]^2  \Big/( {\cal C}_{\rm ex}[p_x p_y,p_x p_y ] + {\cal C}_{\rm c}[p_x p_y,p_x p_y ] ), \label{eqn:visc_microscopic}\\
    \varkappa \geq \frac{1}{4 T^2} \Big[\int \frac{d^2 \bm p}{(2\pi)^2 } n^{(0)}_{\bm p} (1 + n^{(0)}_{\bm p}) p^2 \Big( \frac{p^2}{2m}- \frac{2 T g_2(z)}{g_1(z)} \Big)^2 \Big]^2\Big/\Big({\cal C}_{\rm ex}\big[ p_\alpha p^2,p_\alpha p^2 \big] + {\cal C}_{\rm c}\big[ p_\alpha p^2,p_\alpha p^2 \big]),\label{eqn:kappa_microscopic}
\end{gather}
where the collision brackets are defined as
\begin{align}
    {\cal C}_{\rm ex}[X,Y] & =  \frac{g_{\rm ex}^2}{4} \int  \frac{d^2 \bm p}{(2\pi)^2} \int  \frac{d^2 \bm k}{(2\pi)^2} \int \frac{d^2 \bm p'}{(2\pi)^2} \int \frac{d^2 \bm k'}{(2\pi)^2}   (\bm k\cdot \bm p)^2  (2\pi) \delta( \varepsilon_{p} + \varepsilon_k - \varepsilon_{p'} - \varepsilon_{k'}
 ) \notag\\
    &\times (2\pi)^2\delta(\bm p + \bm k - \bm p' - \bm k') \times  n^{(0)}_{\bm k }n^{(0)}_{\bm p } (1 + n^{(0)}_{\bm k'}) (1 + n^{(0)}_{\bm p'})  {\cal S}[X]{\cal S}[Y],
\end{align}
and
\begin{align}
    {\cal C}_{\rm c}[X,Y] &  = \frac{g_{\rm c}^2}{4} \int  \frac{d^2 \bm p}{(2\pi)^2} \int  \frac{d^2 \bm k}{(2\pi)^2} \int \frac{d^2 \bm p'}{(2\pi)^2} \int \frac{d^2 \bm k'}{(2\pi)^2}    (2\pi) \delta( \varepsilon_{p} + \varepsilon_k - \varepsilon_{p'} - \varepsilon_{k'}
 ) \notag\\
    &\times (2\pi)^2\delta(\bm p + \bm k - \bm p' - \bm k') \times  n^{(0)}_{\bm k }n^{(0)}_{\bm p } (1 + n^{(0)}_{\bm k'}) (1 + n^{(0)}_{\bm p'}) {\cal S}[X] {\cal S}[Y].
\end{align}
Here ${\cal S}[\Phi] = \Phi_{\bm k} + \Phi_{\bm p} - \Phi_{\bm k'} - \Phi_{\bm p'}$. These various multidimensional integrals can be evaluated using the GMM representation in Eq.~\eqref{eq:GMM}, following the derivations in Appendix~\ref{sup:matrix_elements}; see also Ref.~\cite{xue_magnon_2026} for a detailed analysis.

\begin{figure}[!htb]
\centering
\includegraphics[width=0.5\linewidth]{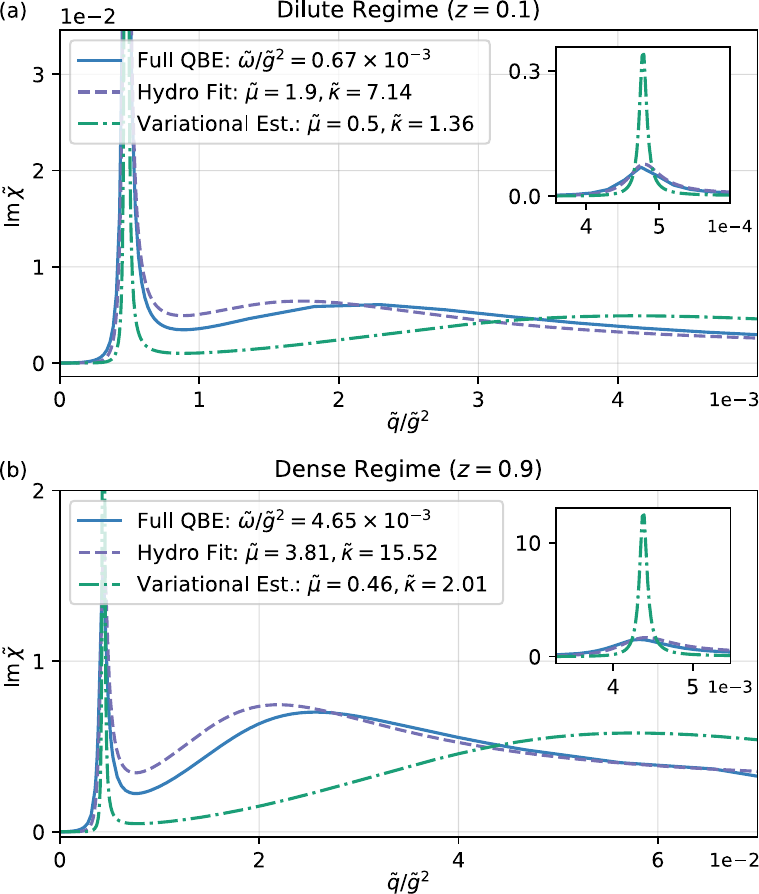} 
\caption{
\textbf{Hydrodynamic approximation to the full QBE solution.}
Representative hydrodynamic fits based on Eq.~\eqref{eq:chi_from_hydro} to constant-frequency cuts of the numerical QBE response for exchange-driven interactions in (a) the dilute regime ($z=0.1$) and (b) the dense regime ($z=0.9$).
The insets highlight the corresponding sound resonances. To determine the effective transport coefficients used to generate the two-dimensional maps in the main text [cf. Fig.~\ref{fig:Combined_2D_hydro_map}], a global fit is performed over a broad $(q,\omega)$-window. This procedure yields $\tilde{\mu}_{\rm ex}\approx1.0,\, \tilde{\varkappa}_{\rm ex}\approx 2.5$ for $z=0.1$ and $\tilde{\mu}_{\rm ex}\approx2.2,\, \tilde{\varkappa}_{\rm ex}\approx 5.4$ for $z=0.9$.
}
\label{fig:Fitted_transport_coeff_exchange}
\end{figure}

\section{Effective Transport Coefficients} \label{sup:fitting_details}

Having established the variational bounds in Appendix~\ref{sec:microscopic_transport_coeff}, we can now compare these estimates with transport coefficients extracted by fitting the hydrodynamic response function in Eq.~\eqref{eq:chi_from_hydro} to the numerical QBE response [Fig.~\ref{fig:Fitted_transport_coeff_exchange}].
Both the full QBE response function and its hydrodynamic counterpart obey the dynamical scaling relation in Eq.~\eqref{eqn:rescale_chi}. As such, the natural scaling variables for these fits are the dimensionless frequency and momentum further rescaled by the coupling constant: $\tilde{\omega}' = \tilde{\omega}/\tilde{g}^2$ and $\tilde{q}' = \tilde{q}/\tilde{g}^2$.

The validity of a hydrodynamic fit depends sensitively on how deeply the system lies within the collective hydrodynamic regime. The hydrodynamic response function~\eqref{eq:chi_from_hydro} is formally applicable only in the long-wavelength, low-frequency limit, $ql_{\rm mfp}\ll1$ and $\omega\tau\ll1$. At finite frequencies and momenta, higher-order gradient corrections in both space and time modify the spectral response. 
In the exchange fluid, these corrections can be particularly significant because the momentum-dependent scattering vertex causes low-momentum magnons to effectively decouple from the hydrodynamic flow.

As a result, transport coefficients extracted from individual constant-frequency slices [\figu{fig:Fitted_transport_coeff_exchange}] exhibit a dependence on the rescaled frequency $\tilde{\omega}'$ (and a weak dependence on the momentum cutoff $\tilde{q}'_{\rm max}$). This dependence reflects the continuous crossover from the hydrodynamic regime at low frequencies to the nonlocal ballistic regime at higher frequencies. 
To obtain the characteristic transport coefficients reported in the main text [\figu{fig:Combined_2D_hydro_map}], we therefore perform a global fit of the hydrodynamic response function~\eqref{eq:chi_from_hydro} to the numerical QBE maps over a broad $(q',\omega')$-window.

In the dilute regime $z=0.1$, this global fit yields $\tilde{\mu}_{\rm ex}\approx1.0$, substantially larger than the variational estimate $\tilde{\mu}_{\rm ex}\approx0.5$. The corresponding variational hydrodynamic response therefore exhibits an artificially narrow and enhanced sound resonance [inset of \figu{fig:Fitted_transport_coeff_exchange}(a)].
In the dense regime $z=0.9$, the fit yields $\tilde{\mu}_{\rm ex}\approx2.2$, more than twice the value extracted in the dilute regime. This increase originates from the accumulation of magnons at low momenta, which, however, participate only weakly in the collective hydrodynamic flow due to their strongly suppressed scattering rates. The variational estimate fails to capture this effect, remaining near $\tilde{\mu}_{\rm ex}\approx0.5$ in both regimes.

\end{document}